\documentclass[12pt]{iopart}

\usepackage{amsmath}
\usepackage{graphicx}
\usepackage{bm}
\usepackage[dvipsnames]{xcolor}
\usepackage{epstopdf}
\usepackage{float}
\usepackage{cite}
\usepackage{multirow}
\usepackage{appendix}
\usepackage{balance}
\usepackage{silence}
\newcommand{\bra}[1]{\langle#1|} 
\newcommand{\ket}[1]{|#1\rangle} 
\newcommand{\normsq}[1]{||#1||^2} 
\usepackage{xparse}

\NewDocumentCommand{\vb}{s m}{%
	\IfBooleanTF{#1}{{\bm{#2}}}{\bm{#2}}%
}
\usepackage{iopams}  
\begin{document}

\title[]{Geometric Discord of any arbitrary dimensional bipartite system and its application in quantum key distribution}

\author{Rashi Jain  \& Satyabrata Adhikari$^*$}

\address{Department of Applied Mathematics,\\ Delhi Technological University, Delhi-110042, India\\
$^*$\begin{footnotesize}Author to whom any correspondence should be addressed.\end{footnotesize}}
\ead{rashijain\_23phdam07@dtu.ac.in, satyabrata@dtu.ac.in }
\vspace{10pt}
\begin{indented}
\item[]October 2025
\end{indented}

	
	
	
	\begin{abstract}
		Entangled quantum states are regarded as a key resource in quantum key distribution (QKD) protocols. However, quantum correlations, other than entanglement can also play a significant role in the QKD protocols. In this work, we will focus on one such measure of quantum correlation, known as geometric quantum discord (GQD). Firstly, we derive an analytical expression of GQD for two-qutrit quantum systems and further generalize it for $d_1\otimes d_2$ dimensional systems. Next, we apply the concept of GQD in studying QKD. In particular, if the shared resource state is an entangled state constructed with the linear combination of the tensor product of the Bell pair and the state $\sigma_i$'s, $i=0,1,2,3$, then we have shown that under some assumption on $\sigma_i$'s, the lower bound for a distillable secret key rate $K_D$ can be expressed in terms of GQD of $\frac{\sigma_0+\sigma_1}{2}$ and $\frac{\sigma_2+\sigma_3}{2}$. Thus, the distillable key rate depends upon the GQD of $\frac{\sigma_0+\sigma_1}{2}$ and $\frac{\sigma_2+\sigma_3}{2}$, when the communicating parties uses private states for generating a secret key in presence of an eavesdropper. Further, for a certain range of GQD of $\frac{\sigma_0+\sigma_1}{2}$ and $\frac{\sigma_2+\sigma_3}{2}$, we find that there exists some NPT entangled resource state for which the successful generation of the secret key may not be guaranteed. We, moreover study the behavior of distillable key rate when the geometric discord of $\frac{\sigma_0+\sigma_1}{2}$ and $\frac{\sigma_2+\sigma_3}{2}$ increases, decreases or remains constant, with the help of a few examples.
	\end{abstract}
	
	\maketitle{}
	\section{Introduction}
	Quantum entangled states served as a resource state and plays a vital role in several tasks, namely, quantum teleportation \cite{bennett_1895, verstraete_2003}, quantum key distribution \cite{gisin_2002} and superdense coding \cite{bennett_1992}. Therefore, it is necessary to know whether a given state is entangled or separable, while performing such tasks. This entanglement detection problem is an important problem in quantum information theory. There are many criterion proposed in the literature for the detection of entangled states. One way of detecting entanglement is by introducing the measure of entanglement. If we are able to quantify the amount of entanglement then we can not only say that the state is entangled but also we can quantify how much the state is entangled simultaneously. The two most important type of entanglement measures are concurrence and negativity. The most preferable measure of entanglement is negativity because of the following two reasons: (i) it is very easy to calculate and (ii) it works in an arbitrary dimensional system. But entanglement is not the only type of quantum correlation, beyond this particular type of quantum correlation, there may exist quantum correlation in a separable state also. Naturally, one may ask a question that how to quantify the quantum correlation present in a separable state as entanglement measure gives zero value for all separable states and therefore it doesn't work in this case. Addressing the issue of the existence of quantum correlation in the composite system, which may be either entangled system or separable system, Oliver and Zurek introduced the notion of quantum discord as a measure of quantum correlations \cite{ollivier_2001}. The idea of quantum discord is to quantify all types of quantum correlations including those not captured by entanglement. Therefore, quantum discord is a measure of quantum correlation that may quantify the quantum correlation in a separable state also.\\
	In the early 2000s, it was found that separable states contain correlations that are not entirely classical \cite{zurek_2000,henderson_2001,modi_2012}. It has been shown that almost all quantum states have non-zero quantum discord \cite{ferraro_2010}. \\
	Mathematically, quantum discord for a bipartite state is defined as follows. Let $H_A$ and $H_B$ be Hilbert spaces associated with systems $A$ and $B$ respectively. The quantum mutual information for the bipartite state $\rho_{AB}\in H_A\otimes H_B$ is given by \cite{ali_2010, girolami_2011}
	\begin{align}
		I(\rho_{AB})=S(\rho_A)+S(\rho_B)-S(\rho_{AB})
		\label{quantum_multual_information}
	\end{align} 
	where $\rho_A$ and $\rho_B$ are the reduced density matrices belong to the Hilbert spaces $H_A$ and $H_B$ respectively, and $S(\rho)=-\text{Tr}(\rho \log_2 \rho)$ is the von Neumann entropy of the state $\rho$. After a measurement on subsystem $A$ under projective operators $\{E_i^A\}$, the conditional state $\rho_i$ is given by \cite{nielsen_2000_book}
	\begin{align}
		\rho_i=\frac{1}{p_i}(E_i^A\otimes I)\rho(E_i^A\otimes I)
	\end{align} 
	with $p_i=\text{Tr}[(E_i^A\otimes I)\rho(E_i^A\otimes I)]$ where $I$ is the identity operator. The classical mutual information, in this case is quantified by
	\begin{align}
		C_A(\rho_{AB})=S(\rho_B)-\min_{\{E_i^A\}}S(\rho_B|\{E_i^A\})
	\end{align}
	The discrepancy between the quantum mutual information and the classical mutual information defines the quantum  discord
	\begin{align}
		D_A(\rho_{AB})=I(\rho_{AB})-C_A(\rho_{AB})
	\end{align} 
	The subscript $A$ in $D_A(\rho_{AB})$ represents that the discord is calculated with respect to the subsystem $A$, and is called as left discord. It is interesting to note here that $D_A(\rho_{AB})$ is always non-negative and asymmetric with respect to subsystems $A$ and $B$ \cite{ollivier_2001}. Due to the complexity of calculating quantum discord of the state in higher dimensional systems, the concept of geometric quantum discord (GQD) has been introduced via the Hilbert-Schmidt norm by Dakic et al. in the following way \cite{dakic_2010}
	\begin{align}
		D_G(\rho_{AB})=\min_{\chi\in C} \normsq{\rho_{AB}-\chi_{AB}}_2=\min_{\chi\in C}\text{Tr}(\rho_{AB}-\chi_{AB})^2
		\label{discord_general}
	\end{align}
	where $C$ denotes the set composed of all zero-discord states, i.e., classical-quantum states, and $\normsq{A}_2=\text{Tr}(A^\dagger A)$, and $\rho_{AB}$ is any two-qubit state
	\begin{align}
		\begin{split}
			\rho_{AB}=&\frac{1}{4}\bigg(I_2\otimes I_2+\sum_{i=1}^{3}x_i \sigma_i\otimes I_2 \sum_{i=1}^{3}y_i I_2 \otimes \sigma_i+\sum_{i,j=1}^{3}T_{ij}\sigma_i\otimes \sigma_j\bigg)
		\end{split}
	\end{align}
	where $\sigma_i$'s are the Pauli matrices and $I_2$ is the identity matrix of order 2. Therefore, the geometric discord of the state $\rho_{AB}$ can be evaluated as
	\begin{align}
		D_G(\rho_{AB})&=\frac{1}{4}(\normsq{\vb*{x}}_2+\normsq{T}_2-\lambda_{max})
	\end{align}
	with $\vb*{x}=(x_1,x_2,x_3)^T$ and $T$ is the correlation matrix corresponding to $\rho$, and $\lambda_{max}$ is the largest eigenvalue of the matrix $\vb*{x}{\vb*{x}}^{T}+TT^T$.
	In particular, geometric discord is a measure, which quantifies the amount of non-classical correlations of a state in terms of its minimum distance from the set of zero-discord states, i.e., classical states.\\
	Quantum discord is a useful concept with various applications in the areas related to quantum information theory, such as quantum teleportation \cite{adhikari_2012}, quantum cryptography \cite{Pirandola_2014}, remote state preparation \cite{pati_2000,dakic_2012}, and super dense coding \cite{shaukat_2022}. Recently, geometric discord has gained a lot of attention due to simpler optimization and easy computation. Most of the work focuses upon the calculation of geometric discord for two-qubit states. In 2010, Luo and Fu calculated the simple and tight lower bound of geometric discord for an arbitrary state \cite{luo_2010}. Later, Wang et al. provided a lower bound, which is an approximation of GQD of the given two-qubit state \cite{wang_2024}.  Nevertheless, an exact analytical formula for geometric discord was put forward recently, in 2024, by Loubenets and Hanotel for arbitrary two qudits states \cite{loubenets_2024}. Various analytical formula of geometric discord for multipartite systems have already been proposed \cite{radhakrishnan_2020,zhu_2022,hassan_2024}. \\
	On the practical aspects of QKD, including the measurement-free mediated semi-quantum protocols \cite{zhou_2024}, implementation security \cite{zapatero_2025}, measurement device-independent QKD \cite{trenyi_2019}, and real-time microsatellite-based QKD \cite{li_2025}, which demonstrate the growing relevance of both experimental and theoretical advances in the field.  In addition to these, studies on noisy quantum channels have focused upon the critical impact of environmental noise on QKD \cite{teklu_2015, trapani_2015, adnane_2019, ikuta_2016}. Recently, it has been found that quantum discord plays a crucial role in quantum cryptography, specifically in quantum key distribution (QKD) \cite{Pirandola_2014}. In 2023, Lai et al \cite{lai_2023} pointed out the importance of quantum correlations in ensuring secure communication.  Discord-based QKD protocols have also gained a lot of attention \cite{zhu_2022, wang_2024}. Gu et al \cite{gu_2017} introduced the trusted noise QKD. Bera et al \cite{bera_2017} discussed the applications of quantum discord in the QKD. These results suggests that even in the absence of entanglement, quantum correlations like discord can still enable secure key distribution. However, despite all these studies, we observe that there is a gap where the role of GQD, has remained unnoticed in context of QKD. Thus, in this work, we will focus on the scenario where secure key distribution is possible irrespective of whether the entanglement is present or absent between the sender and the receiver. To do this, we make use of $2\otimes 2$ and $3\otimes 3$ dimensional private states, which securely carries the key part of the shared quantum state, and is responsible for the secure secret key generation \cite{Horodecki_2008,Horodecki_2005_prl_94}. \\
	The aim of this study is two-fold. Firstly, we are interested in characterizing the GQD in case of two-qudit states. To do this, we derive an analytical expression of geometric discord for two-qutrit states and further generalize it for two-qudit systems. Using the analytical expression obtained, we quantify the quantum correlation present in a higher dimensional separable states, negative partial transpose entangled states (NPTES) and positive partial transpose entangled states (PPTES). We also study the relationship between negativity and geometric discord. Secondly, we establish a new lower bound in terms of GQD on the distillable key rate. This will help in studying QKD protocols even with the separable states with non-zero GQD. Therefore, if the distillable key rate will be positive even with the separable state having non-zero discord then this fact indicates the success of the key distribution protocol and in this way, we will be able to connect geometric discord directly to the privacy in quantum communication.\\
	The rest of the paper is organized as follows. In section II, we derive the analytic expression of GQD for a two-qutrit system, and propose a generalized formula for a two-qudit system. We use our formula to calculate the geometric discord of separable states, NPT entangled states and PPT entangled states. In section III, we demonstrate a few examples to show whether there exist any possible relationship between the GQD and negativity. In section IV, we discuss the applications of GQD in quantum key distribution by deriving a lower bound of the distillable key rate in terms of GQD. Finally, we conclude in section V.

	\section{Generalized geometric discord of a bipartite  system}
	Brukner et.al. derived the geometric discord for two-qubit system and in this section, we have adopted their strategy to derive the analytic expression of geometric discord for a two-qutrit system and then generalize it to the two-qudit system. Then we discuss a few examples where we used our derived formula to calculate the geometric discord of the separable states, the NPT entangled states (free entangled states) and the PPT entangled states (bound entangled states) respectively.
	\subsection{Analytic expression for geometric discord of $d_{1}\otimes d_{2} $ system}
	To start with, we will first derive the expression of geometric discord of $3\otimes 3$ system and then find the expression of the geometric discord of $d_{1} \otimes d_{2}$ dimensional system. Let us consider the two-qutrit bipartite system described by the Hilbert spaces $\mathbb{H}_A$ and $\mathbb{H}_B$ respectively. Then a generalized two-qutrit state $\rho^{3\otimes 3}$ can be expressed in a Bloch representation as \cite{ferraro_2010}
	
	\begin{align}
		\begin{split}
			\rho_{AB}^{3\otimes 3}=&\frac{1}{9}\bigg(I_3 \otimes I_3 +\sum_{i=1}^{8} x_i \Lambda^{(i)}\otimes I_3 + \sum_{i=1}^{8} y_i I_3 \otimes \Lambda^{(i)} +\sum_{i,j=1}^{8}T_{ij} \Lambda^{(i)}\otimes \Lambda ^{(j)}\bigg)
		\end{split}
		\label{twoqutrit_our}
	\end{align}
	where $I_3$ denote the identity matrix of order 3 and $\Lambda^{(i)}, i\in \{1,2,\dots ,8\}$ represent the eight Gell-Mann matrices defined as follows \cite{bertlmann_2008}
		\begin{small}
			\begin{align}
				\begin{split}
					&\ket{j}\bra{k}+\ket{k}\bra{j}, ~1\leq j<k \leq 3,~i=1,2,3  \\
					&\ket{j}\bra{k}+\ket{k}\bra{j}, ~1\leq j<k \leq 3,~  i=4,5,6\\
					&\sqrt{\frac{2}{l(l+1)}} \left(\sum_{j=1}^{l}\ket{j}\bra{j}-l\ket{l+1}\bra{l+1}\right), ~1\leq l \leq 2,~ i=7,8
					\label{gell_mann matrices}
				\end{split}
			\end{align}
		\end{small}
	The properties of the above defined Gell-Mann matrices are as follows:\\
	(i) They are traceless, i.e., $\text{Tr}[\Lambda^{(i)}]=0$,~~for $i=1...8$.\\
	(ii) They are orthogonal i.e., for $i\neq j$,~ $\Lambda^{(i)} \cdot \Lambda^{(j)}=0$, for $i,j=1,2,\dots,8$.\\ 
	(iii) $\text{Tr}[(\Lambda^{(i)})^{2}]=2$, for $i=1,2,\dots,8$.\\
	The coefficients $x_{i}$, $y_{i}$ and $T_{ij}$ can be calculated as 
	\begin{align}
		\begin{split}
			x_i=&\text{Tr}(\rho(\Lambda^{(i)}\otimes I_3)),~~i=1...8\\
			y_i=&\text{Tr}(\rho(I_3 \otimes \Lambda^{(i)})),~~ i=1...8\\
			T_{ij}=&\text{Tr}(\rho (\Lambda^{(i)}\otimes \Lambda^{(j)})),~~ i,j=1...8
		\end{split}
	\end{align} 
	If the coefficients $x_{i}$, $y_{i}$ denote the components of the Bloch vectors $\vb*{x}$, $\vb*{y}$ and $T_{ij}$ denotes the components of the correlation tensor then the two-qutrit state $\rho$ can be expressed in the form of the triplet as $\rho=\{\vb*{x},\vb*{y}, T\}$.\\
	Let $C$ be the set of two-qutrit zero-discord states, i.e., the set $C$ represent the set of two-qutrit classical-quantum states $\chi_{AB}^{(3\otimes3)}$, which can be expressed in the form as 
	\begin{align}
		\chi_{AB}^{(3\otimes 3)}=p_1 \ket{\psi_1}\bra{\psi_1} \otimes \rho_1 +p_2 \ket{\psi_2}\bra{\psi_2} \otimes \rho_2 + p_3 \ket{\psi_3}\bra{\psi_3} \otimes \rho_3
	\end{align}
	where $p_1,~p_2$ and $p_3$ are non-negative numbers, such that $p_1+p_2+p_3=1$ and $\rho_k, ~ k=1,2,3$ denotes the single qutrit generalized density matrices, given by
	\begin{align}
		\begin{split}
			\rho_1=\frac{1}{3} I_3 +\sum_{j=1}^{8} b_{j} \Lambda^{(j)} ,~ |\vb*{b}|\leq \sqrt{\frac{1}{3}}\\
			\rho_2=\frac{1}{3} I_3 +\sum_{j=1}^{8} c_{j} \Lambda^{(j)} ,~ |\vb*{c}|\leq \sqrt{\frac{1}{3}}\\
			\rho_3=\frac{1}{3} I_3 +\sum_{j=1}^{8} d_{j} \Lambda^{(j)} ,~ |\vb*{d}|\leq \sqrt{\frac{1}{3}}
		\end{split}
		\label{rho_k}
	\end{align}
	where $\vb*{b},\vb*{c}$ and $\vb*{d}$ are column vectors with each component $b_j=\frac{1}{2}\text{Tr}[\rho_1 \Lambda^{(j)}]$, $c_j=\frac{1}{2}\text{Tr}[\rho_2 \Lambda^{(j)}]$ and $d_j=\frac{1}{2}\text{Tr}[\rho_3 \Lambda^{(j)}]$ respectively.\\
	The single-qubit states $\{\ket{\psi_1},\ket{\psi_2},\ket{\psi_3}\}$ forms an orthonormal basis, which satisfies the completeness relation i.e. $\ket{\psi_1}\bra{\psi_1}+\ket{\psi_2}\bra{\psi_2}+\ket{\psi_3}\bra{\psi_3}=I_3$. Let us choose the normalized states $\{\ket{\psi_1},\ket{\psi_2},\ket{\psi_3}\}$ as follows
	\begin{align}
		\begin{split}
			\ket{\psi_1}&=\alpha \ket{0}+\beta \ket{1}+\gamma \ket{2}\\
			\ket{\psi_2}&=\frac{-\beta \ket{0}+\alpha \ket{1}}{\sqrt{\alpha^2+\beta^2}}\\
			\ket{\psi_3}&=\frac{-\alpha \gamma \ket{0}-\beta \gamma \ket{1}+(\alpha^2+\beta^2)\ket{2}}{\sqrt{\alpha^2+\beta^2}}
		\end{split}
		\label{psi}
	\end{align}
	where $\alpha,\beta,\gamma\in \mathbb{R}$ satisfy the normalization condition $\alpha^2+\beta^2+\gamma^2=1$.\\
	In terms of triplet, the zero-discord state $\chi_{AB}^{3\otimes 3}$ can be expressed as $\chi=\{a_1^{(i)}t_1+a_2^{(i)}t_2, \vb*{u}, S\}$. The first term of the triplet is given by
	\begin{eqnarray}
		a_1^{(i)}t_1+a_2^{(i)}t_2=\sum_{k=1}^{2}a_{k}^{(i)}t_{k},~i=1,2,\dots,8
	\end{eqnarray}
	where $a_k^{(i)}=\bra{\psi_k}\Lambda^{(i)}\ket{\psi_k}$,
	for $k=1,2$, and $t_1=p_1-p_3$, $t_2=p_2-p_3$ (Refer to Appendix A.1 for detailed calculations).\\
	The second term of the triplet is given by the eight component vector $\vb*{u}$ which can be written as (Refer to Appendix A.2 for detailed calculations)
	\begin{align}
		\begin{split}
			\vb*{u}=   \bigg(&\text{Tr}\left[\left(\sum_{k=1}^{3}p_k \rho_k\right)\Lambda^{(1)}\right],\text{Tr}\left[\left(\sum_{k=1}^{3}p_k\rho_k\right)\Lambda^{(2)}\right],\dots, \text{Tr}\left[\left(\sum_{k=1}^{3}p_k\rho_k\right)\Lambda^{(8)}\right]\bigg)
		\end{split}    
	\end{align}
	The elements of the third term of the triplet, $S$ are given by
	\begin{align}
		S^{(i,j)}=\vb*{a^{(i)}}\cdot \vb*{r^{(j)}}=\sum_{k=1}^{2}a_{k}^{(i)}r_{k}^{(j)},~ i,j=1...8
	\end{align}
	where $r_{1}^{(j)}$ and $r_{2}^{(j)}$ are given by 
	\begin{eqnarray}
		&&r_{1}^{(j)}=\text{Tr}[(p_1\rho_1-p_3\rho_3)\Lambda^{(j)}],j=1...8\nonumber\\&&
		r_{2}^{(j)}=\text{Tr}[(p_2\rho_2-p_3\rho_3)\Lambda^{(j)}],j=1...8
	\end{eqnarray}
	It can be observed here that $S^{(i,j)}, i,j=1,2,\dots, 8$ denotes the matrix elements of a $8\times 8$ correlation matrix $S$ of the zero-discord state $\chi_{AB}^{3\otimes 3}$ (Refer to Appendix A.3 for detailed calculations).\\	
	Additionally, the components to the Bloch vectors of the state $\chi_{AB}^{3\otimes 3}$ have $\normsq{\vb*{a^{(i)}}}_2=\frac{4}{3}$, ~$\normsq{\vb*{u}}_2\leq\frac{1}{3}$,~ and $\normsq{S}_2\leq\frac{4}{9}$,~$k=1,2,3$,~$i=1,2,\dots, 8$.\\
	Now, we are in a position to calculate the geometric discord of a bipartite $3 \otimes 3$ system. Geometric discord of a bipartite $3 \otimes 3$ system described by the density operator $\rho_{AB}^{3\otimes 3}$ is defined as the  minimum Hilbert Schmidt distance between $\rho_{AB}^{3\otimes 3}$ and the zero discord state $\chi_{AB}^{3\otimes 3}$, which is given by \cite{witte_1999,left_discord}
	\begin{align}
		\begin{split}
			D^{(3,3)}_G(\rho_{AB}^{3\otimes 3})&=\min_{\chi_{AB}^{3\otimes 3}}\normsq{\rho_{AB}^{3\otimes 3}-\chi_{AB}^{3\otimes 3}}_2=\min_{\chi_{AB}^{3\otimes 3}} \text{Tr}(\rho_{AB}^{3\otimes 3}-\chi_{AB}^{3\otimes 3})^2\\
			&=\min_{\chi_{AB}^{3\otimes 3}}(\normsq{\rho_{AB}^{3\otimes 3}}_2+\normsq{\chi_{AB}^{3\otimes 3}}_2-2\text{Tr}(\rho_{AB}^{3\otimes 3} \chi_{AB}^{3\otimes 3}))
		\end{split}
		\label{norm_rho-chi_formula}
	\end{align}
	Now, substituting the value of $\normsq{\rho_{AB}^{3\otimes 3}}_2$ and $\normsq{\chi_{AB}^{3\otimes 3}}_2$ in terms of the triplets $\{\vb*{x},\vb*{y},T\}$ and $\{a_1^{(i)}t_1+a_2^{(i)}t_2, \vb*{u}, S\}$, (\ref{norm_rho-chi_formula}) can be re-expressed as (Refer to Appendix B for detailed calculations)
	\begin{align}
		\begin{split}
			D^{(3,3)}_G(\rho_{AB}^{3\otimes 3})&=\min\bigg[\frac{2}{81} \bigg(3 \normsq{\vb*{x}}_2+3\normsq{\vb*{y}}_2+2 \normsq{T}+4(t_1^2+t_2^2-t_1t_2)+3 \normsq{\vb*{u}}_2+2 \normsq{\vb*{S}}_2\\
			&-6 (t_1{\vb*{x}}^T \cdot\vb{a_1^{(i)}}+t_2{\vb*{x}}^T \cdot\vb{a_2^{(i)}}) -6 \vb*{y}\cdot \vb*{u}-4 ({\vb*{r_1^{(j)}}}^T T \vb*{a_1^{(i)}}+{\vb*{r_2^{(j)}}}^T T \vb*{a_2^{(i)}})\bigg)\bigg]
		\end{split}
		\label{norm_rho_chi}
	\end{align}
	where the column vectors $\vb*{r_{1}^{(j)}}$ and $\vb*{r_{2}^{(j)}}$ are given by
	\begin{align}
		\vb*{r_1^{(j)}}=\begin{pmatrix}
			r_1^{(1)}\\
			r_1^{(2)}\\
			r_1^{(3)}\\
			r_1^{(4)}\\
			r_1^{(5)}\\
			r_1^{(6)}\\
			r_1^{(7)}\\
			r_1^{(8)}
		\end{pmatrix}, &~~ \vb*{r_2^{(j)}}=\begin{pmatrix}
			r_2^{(1)}\\
			r_2^{(2)}\\
			r_2^{(3)}\\
			r_2^{(4)}\\
			r_2^{(5)}\\
			r_2^{(6)}\\
			r_2^{(7)}\\
			r_2^{(8)}
		\end{pmatrix}
	\end{align}
	and the minimum is taken over the parameters of the zero discord state $\chi_{AB}^{3\otimes 3}$ i.e., $\vb*{a_k^{(i)}}, \vb*{t}, \vb*{u}$ and $\vb*{r_k^{(j)}}$, $k=1,2$.\\
	The minimum of the expression (\ref{norm_rho_chi}) can be obtained for the values of the parameters, which is given as 
	\begin{align}
		\begin{split}
			t_1=&{\vb*{x}}^T\vb{a_1^{(i)}}+\frac{1}{2}{\vb*{x}}^T\vb{a_2^{(i)}}\\
			t_2=&\frac{1}{2}{\vb*{x}}^T\vb{a_1^{(i)}}+{\vb*{x}}^T\vb{a_2^{(i)}}\\
			\vb*{u}=&\vb*{y}\\
			\vb*{r_1^{(j)}}=&\frac{3}{4} T\left(\vb*{a_1^{(i)}}\left({\vb*{a_2^{(i)}}}^T\vb{a_2^{(i)}}\right)+\frac{2}{3}\vb*{a_2^{(i)}}\right)\\
			\vb*{r_2^{(j)}}=&\frac{3}{4} T\left(\vb*{a_2^{(i)}}\left({\vb*{a_1^{(i)}}}^T\vb{a_1^{(i)}}\right)+\frac{2}{3}\vb*{a_1^{(i)}}\right)
		\end{split}
		\label{minimum_sol}
	\end{align}
	The solution obtained in (\ref{minimum_sol}) can be found in Appendix C. Using (\ref{minimum_sol}), the minimum value of the expression given in (\ref{norm_rho_chi}) can be obtained as
	\begin{align}
		\begin{split}
			D^{(3,3)}_G(\rho_{AB}^{3\otimes 3})&=\min_{\vb*{a_1^{(i)}},\vb*{a_2^{(i)}}}\frac{2}{81}\big(3\normsq{\vb*{x}}_2+2\normsq{T}_2\big)\\
			&-\frac{2}{81}\bigg[{\vb*{a_1^{(i)}}}^TM\vb{a_1^{(i)}}+{\vb*{a_2^{(i)}}}^TM\vb{a_2^{(i)}}\bigg]-\frac{1}{81}\bigg[{\vb*{a_1^{(i)}}}^TM\vb{a_2^{(i)}}+{\vb*{a_2^{(i)}}}^TM\vb{a_1^{(i)}}\bigg]
		\end{split}
		\label{norm_rho_chi_min}
	\end{align}
	where $M=3\vb{x}{\vb*{x}}^T+2T^T T$, which denotes the $8\times 8$ matrix having eight eigenvalues $\lambda_1,\lambda_2\dots \lambda_8$, taken in order $\lambda_1\geq\lambda_2\geq \dots \geq\lambda_8$. The two largest eigenvalue of $M$ are denoted by $\lambda_{max}^{(1)}$ and $\lambda_{max}^{(2)}$ respectively. We find that
	\begin{align}
		\begin{split}
			{\vb*{a_k^{(i)}}}^TM \vb*{a_k^{(i)}}=&\frac{4}{3}\lambda_{max}^{(k)}, ~k=1,2\\
			{\vb*{a_k^{(i)}}}^TM \vb*{a_l^{(i)}}=&-\frac{2}{3}\lambda_{max}^{(l)}, ~k\neq l, ~k,l=1,2
		\end{split}
		\label{lambda_max terms}
	\end{align}
	Substituting (\ref{lambda_max terms}) in (\ref{norm_rho_chi_min}) and simplifying, we get
	\begin{align}
		\begin{split}
			D^{(3,3)}_G(\rho_{AB}^{3\otimes 3})=&\frac{2}{81}\bigg[3\normsq{\vb*{x}}_2+2\normsq{T}_2-(\lambda_{max}^{(1)}+\lambda_{max}^{(2)})\bigg]
		\end{split}
		\label{discord_formula_our}
	\end{align}
	Therefore, (\ref{discord_formula_our}) gives the formula for the discord of any $3\otimes 3$ dimensional quantum state described by the density operator $\rho_{AB}^{3\otimes 3}$.\\
	Now our task is to extend this idea for deriving the analytical expression of geometric discord to $d_1\otimes d_2$ dimensional system. To do this, let us consider the two-qudit state  $\rho_{AB}^{d_1\otimes d_2}$ in $d_{1}\otimes d_{2}$ dimensional system. It can be expressed as \cite{ferraro_2010}
	\begin{align}
		\begin{split}
			\rho_{AB}^{d_1\otimes d_2}=&\frac{1}{d_1 d_2}\bigg[I_{d_1}\otimes I_{d_2}+ \sum_{i=1}^{d_1^2-1}x_i \Lambda^{(i)}\otimes I_{d_2}+I_{d_1} \otimes \sum_{i=1}^{d_2^2-1}\Lambda^{(i)}+ \sum_{i=1}^{d_1^2-1}\sum_{j=1}^{d_2^2-1}T(\Lambda^{(i)}\otimes\Lambda^{(j)})\bigg] 
		\end{split}
		\label{rho_two_qudit}
	\end{align}
	where $I_{d_1}$ and $I_{d_2}$ are the identity matrices of order $d_1$ and $d_2$ respectively. $\Lambda^{(i)}$ represents the traceless operators. The coefficients $x_i$ and $y_i$ are the Bloch vectors corresponding to the first and the second qubit respectively, and $T_{ij}$ denotes the correlation matrix corresponding to the state $\rho_{AB}^{d_1\otimes d_2}$.\\
	Following the similar procedure of calculating the geometric discord of the state in $3\otimes 3$ dimensional system, the geometric discord of the state $\rho_{AB}^{d_1\otimes d_2}$ can be calculated as 
	\begin{align}
		\begin{split}
			D_G^{(d_1,d_2)}(\rho_{AB}^{d_1\otimes d_2})=& \frac{2}{d_1^2 d_2} \normsq{\vb*{x}}_2+\frac{4}{d_1^2 d_2^2} \normsq{T}_2-(\lambda_1+\lambda_2+\dots+\lambda_{d_1-1})
		\end{split}
		\label{gd_d1d2}
	\end{align}
	where $\lambda_1\geq \lambda_2...\geq \lambda_{d_1-1}\geq \lambda_{d_1}$ denotes the eigenvalues of the matrix $\frac{2}{d_1^2 d_2} \vb*{x}{\vb*{x}}^T+\frac{4}{d_1^2 d_2^2}T^TT$ of order $(d_1^2-1)\times (d_2^2-1)$.\\

	\subsection{Examples}
	In this section, we will consider a few examples of  product state, diagonal state, separable states, NPT entangled states and PPT entangled states in $3\otimes 3$ and $4\otimes 4$ dimensional system and calculate their geometric discord.
	
	\subsubsection{Geometric discord for product state}
	
    \textbf{Example 1:} Let us first examine the general form of a two-qutrit product state described by the density operator $\rho_{prod}=\rho_1\otimes\rho_2$, where $\rho_1=\frac{1}{3}I_3+\vb*{b}\cdot \vb*{\Lambda}$ and $\rho_2=\frac{1}{3}I_3+\vb*{c}\cdot \vb*{\Lambda}$, where $\vb*{b}=(b_1,b_2\dots, b_8)^T$, $\vb*{c}=(c_1,c_2\dots, c_8)^T$ and $\vb*{\Lambda}$ represents the eight Gell-Mann matrices given in (\ref{gell_mann matrices}). Therefore, we have 
	\begin{align}
		\begin{split}
			\rho_{prod}=&\frac{1}{9} I_3\otimes I_3 +\frac{1}{3} I_3 \otimes \vb*{c}\cdot \vb*{\Lambda}+\frac{1}{3}\vb*{b}\cdot\vb*{\Lambda}\otimes I_3 +\vb*{b}\cdot\vb*{\Lambda}\otimes \vb*{c}\cdot \vb*{\Lambda}
		\end{split}
	\end{align}
	Using the formula of geometric discord (\ref{discord_formula_our}), we calculate the geometric discord of the product state $\rho_{prod}$ and find that 
	\begin{align}
		D^{(3,3)}_G(\rho_{prod})=0
	\end{align}
	This establishes the well known fact that there does not exists any correlations in a product state.\\
	\subsubsection{Geometric discord of separable states}
    \textbf{Example 1:} Let us now consider a general two-qutrit diagonal state $\rho_{diag}$ given by
	\begin{align}
		\rho_{diag}=\sum_{i,j=0}^{2}a_{ij}\ket{ij}\bra{ij}
		\label{rho_diag}
	\end{align}
	The geometric discord of the diagonal state $\rho_{diag}$ is given by
	\begin{align}
		D^{(3,3)}_G(\rho_{diag})=0
	\end{align}
	Thus, a general two-qutrit diagonal state which is a separable state with zero discord exhibits no quantum correlations.\\
	\textbf{Example 2:} Consider a family of two-qutrit isotropic states that are invariant under any unitary transformation $U$ of the form $U\otimes U^*$ \cite{horodecki_1999_1}. It is described by the density operator \cite{rudolph_2003}
	\begin{align}
		\rho_{\beta}=\beta\ket{\phi^+}\bra{\phi^+}+\frac{1-\beta}{9} I_9, ~ -\frac{1}{8}\leq \beta \leq \frac{1}{3}
		\label{sep_rho_beta}
	\end{align}
	where $I_9$ represents the $9\times 9$ identity matrix and $\ket{\phi^+}$ represents the two-qutrit Bell-state, expressed as $\ket{\phi^+}=\frac{1}{\sqrt{3}}(\ket{00}+\ket{11}+\ket{22})$. The state $\rho_{\beta}$ represents a separable state when $-\frac{1}{8}\leq \beta \leq \frac{1}{3}$ and its geometric discord is given by
	\begin{align}
		D_G^{(3,3)}(\rho_{\beta})=\frac{32}{243} \beta^2
		\label{gd_sep_rho_isotropic}
	\end{align}
	\begin{figure}[ht]
		\centering
		\includegraphics[width=0.7\linewidth]{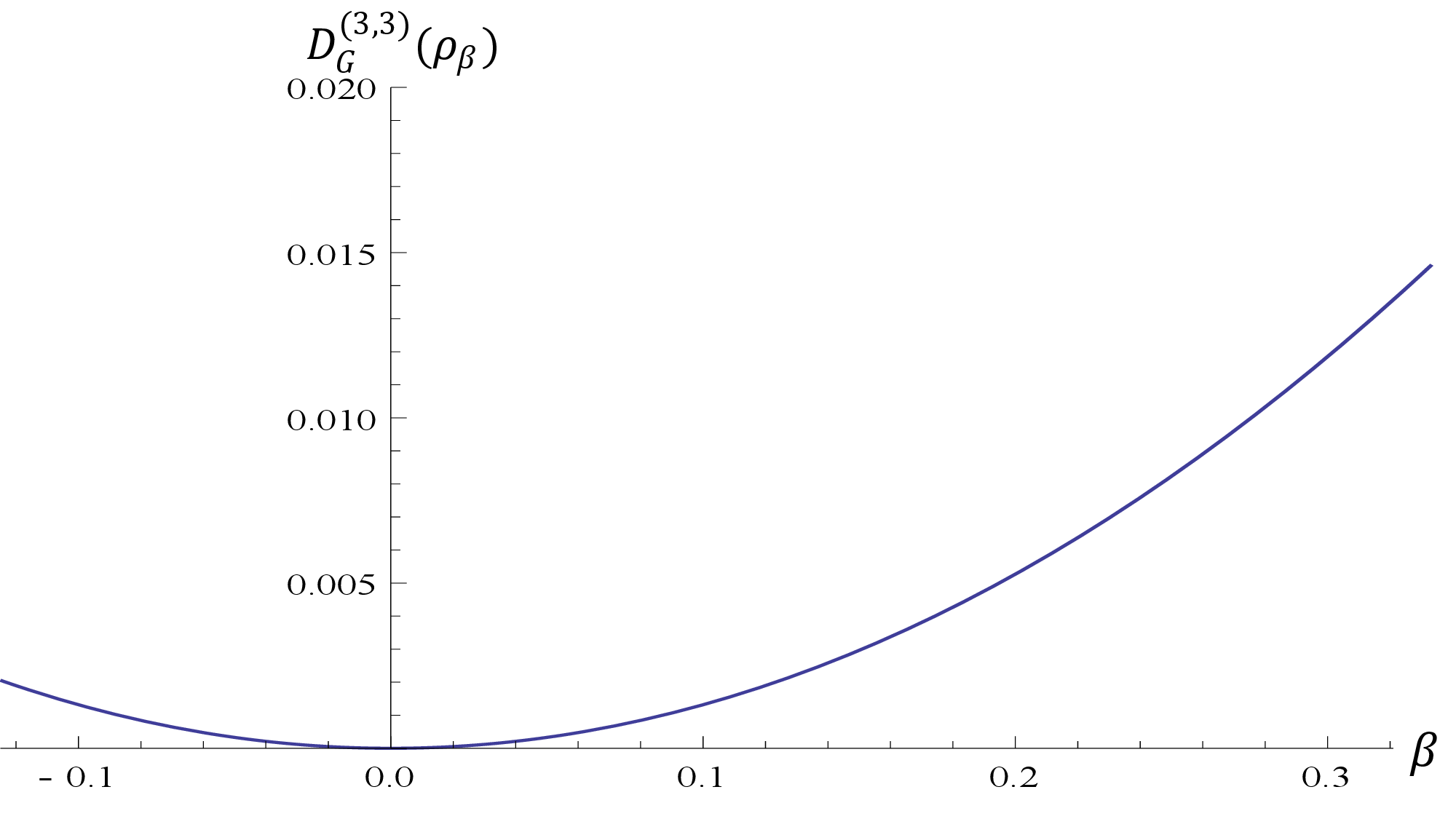}
		\caption{Geometric discord for the isotropic separable state $\rho_{\beta}$, where $\frac{-1}{8}\leq \beta \leq \frac{1}{3}$. The $X$ axis represents the value of the state parameter $\beta$ lying between $\frac{-1}{8}$ and $\frac{1}{3}$, and the $Y$ axis represents the corresponding geometric discord $D_G^{(3,3)}(\rho_{\beta})$ given by (\ref{gd_sep_rho_isotropic}).}
		\label{fig:1our_sep_isotropic}
	\end{figure}
	From (\ref{gd_sep_rho_isotropic}), it can be seen that the geometric discord of the separable state described by the density operator $\rho_{\beta}$ is non-vanishing unless $\beta=0$. Also, it can be observed from FIG. \ref{fig:1our_sep_isotropic} that as the state parameter $\beta$ varies from $-\frac{1}{8}$ to $0$, the geometric discord $D_G^{(3,3)}(\rho_{\beta})$ decreases, and then as the state parameter increases from $0$ to $\frac{1}{3}$, the geometric discord increases. At $\beta=0$, the state $\rho_{\beta}$ reduces to maximally mixed state, i.e., $I_9$ which is also a diagonal state and thus, when $\beta=0$, the geometric discord reduces to zero.\\
	\textbf{Example 3:} A $3\otimes 3$ dimensional separable quantum state described by the density operator $\rho_{\alpha}$ is given by \cite{horodecki_1999}
	\begin{align}
		\rho_{\alpha}=&\frac{2}{7}\ket{\psi^+}\bra{\psi^+}+\frac{\alpha}{7}\sigma^++\frac{5-\alpha}{7}\sigma^-,~ 2\leq \alpha\leq 3
		\label{alpha_state}
	\end{align}
	where $\ket{\psi^+},~\sigma^+$ and $\sigma^-$ are given by 
	\begin{align}
		\begin{split}
			\ket{\psi^+}=&\frac{1}{\sqrt{3}} (\ket{00}+\ket{11}+\ket{22})\\
			\sigma^+=&\frac{1}{3}(\ket{01}\bra{01}+\ket{12}\bra{12}+\ket{20}\bra{20})\\
			\sigma^-=&\frac{1}{3}(\ket{10}\bra{10}+\ket{21}\bra{21}+\ket{02}\bra{02})
		\end{split}
	\end{align}
	
	The geometric discord for the separable state $\rho_{\alpha}$ is given by (Refer to Appendix D for detailed calculations (\ref{our_separable_alpha_d_proof}))
	\begin{align}
		D_G^{(3,3)}(\rho_{\alpha})=\frac{32}{11907}\bigg[\left(\alpha-\frac{5}{2}\right)^2+\frac{11}{4}\bigg],~ 2\leq \alpha \leq 3
		\label{gd_sep_rho_alpha}
	\end{align}
	
	\begin{figure}[ht]
		\centering
		\includegraphics[width=0.7\linewidth]{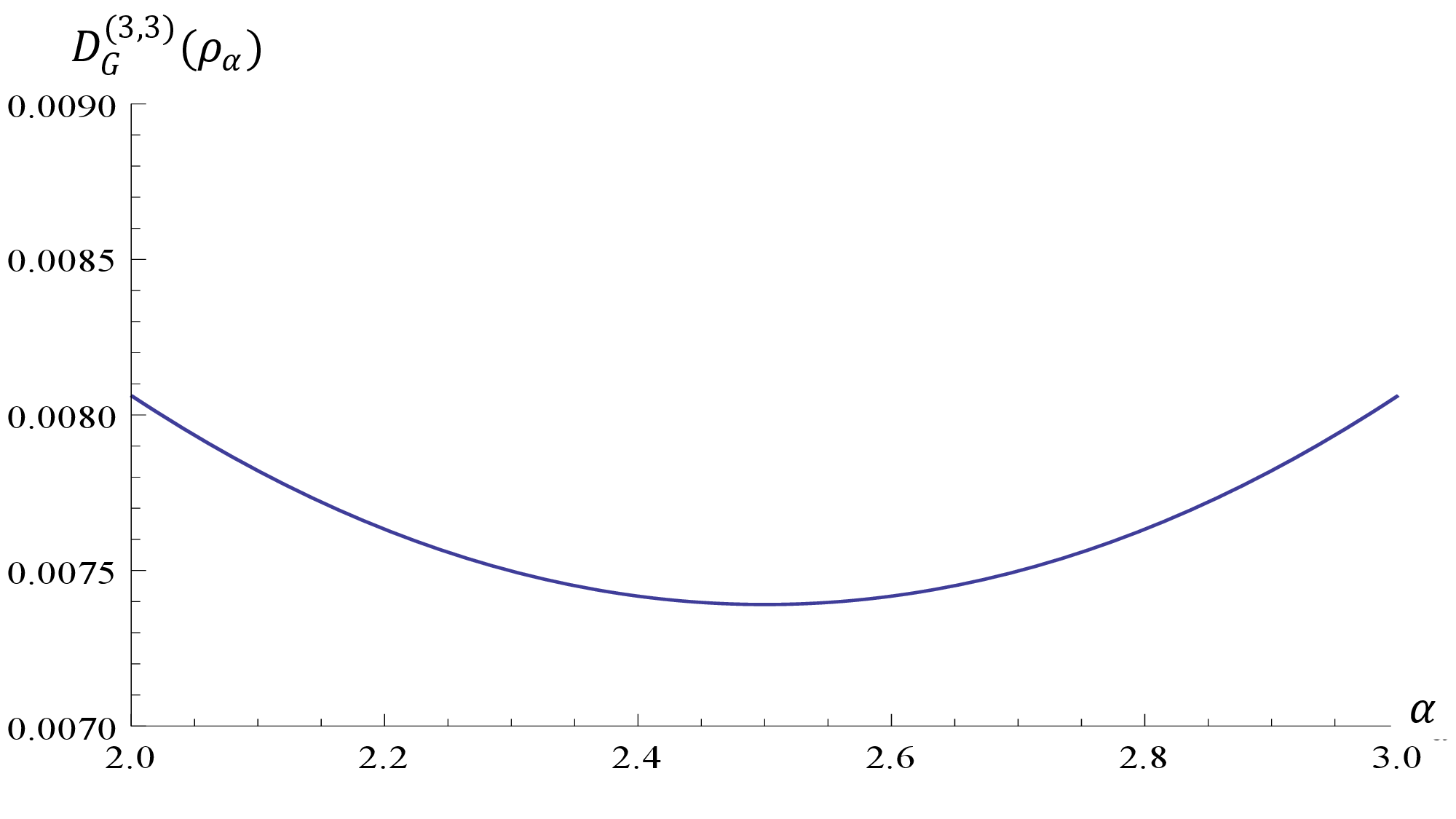}
		\caption{Geometric discord for the state $\rho_{\alpha}$, where $2\leq \alpha \leq 3$ is plotted. The $X$ axis represents the value of the state parameter $\alpha$ lying in the interval $[2,3]$, and the $Y$ axis represents the corresponding geometric discord $D_G^{(3,3)}(\rho_{\alpha})$ given by (\ref{gd_sep_rho_alpha}).}
		\label{fig:1our_sep_alpha}
	\end{figure}
	From (\ref{gd_sep_rho_alpha}), we can easily say that the geometric discord is non-zero for all $\alpha \in [2,3]$. Therefore, the state $\rho_{\alpha}$ is a separable state with non-zero discord and thus it possesses quantum correlation when $2\leq \alpha \leq 3$. Also, it can be observed from FIG. \ref{fig:1our_sep_alpha} that the geometric discord shows a symmetric behavior about $\alpha=2.5$, in the range $2\leq \alpha \leq 3$.\\
	
	\textbf{Example 4:} Let us consider the two-qutrit separable state, given by \cite{bandhyopadhyay_2005}
	\begin{align}
		\rho_1(\gamma)=\gamma \ket{\psi_1}\bra{\psi_1}+(1-\gamma)\rho_{\psi},~ \frac{1}{5}\leq \gamma \leq 1
		\label{sep_rhogamma}
	\end{align}
	where 
	\begin{align}
		\rho_{\psi}=\frac{1}{4}\left(I_9-\sum_{i=1}^{5}\ket{\psi_i}\bra{\psi_i}\right)
		\label{rho_psi}
	\end{align}
	and $\ket{\psi_i}$ for $i=1,2,\cdots, 5$ are defined as
	\begin{align}
		\begin{split}
			\ket{\psi_1}=&\frac{1}{\sqrt{2}}\ket{0}\otimes(\ket{0}-\ket{1})\\
			\ket{\psi_2}=&\frac{1}{\sqrt{2}}(\ket{0}-\ket{1})\otimes\ket{2}\\
			\ket{\psi_3}=&\frac{1}{\sqrt{2}}\ket{2}\otimes (\ket{1}-\ket{2})\\
			\ket{\psi_4}=&\frac{1}{\sqrt{2}}(\ket{1}-\ket{2})\otimes\ket{0}\\
			\ket{\psi_5}=&\frac{1}{3}(\ket{0}+\ket{1}+\ket{2})\otimes (\ket{0}+\ket{1}+\ket{2})
		\end{split}
		\label{rho_psi_i}
	\end{align}
	The geometric discord of $\rho_1(\gamma)$ is given by
	\begin{align}
		D_G^{(3,3)}(\rho_1(\gamma))=\frac{2}{81}\bigg[&\frac{1}{225} (611 - 1472 \gamma+ 2561 \gamma^2)-\lambda_{max}^{(1)}+\lambda_{max}^{(2)})\bigg]
		\label{gd_sep_rhogammai}
	\end{align}
	where $\lambda_{max}^{(1)}$ and $\lambda_{max}^{(2)}$ are the two largest root of $-1024 (-1 + \gamma)^6 (73 + \gamma (-878 + \gamma (4051 - 8760 \gamma + 8514 g^2))) + 
	160 (-1 + \gamma)^4 (21578 + 
	\gamma (-198782 + \gamma (863353 + \gamma (-2821672 + 5743023 \gamma)))) x - 
	500 (-1 + \gamma)^2 (94076 + 
	\gamma (-734364 + \gamma (3167711 + \gamma (-6862134 + 9869711 \gamma)))) x^2 + 
	93750 (2582 + \gamma (-15648 + \gamma (55577 + \gamma (-83538 + 63527 \gamma)))) x^3 - 
	703125 (611 + \gamma (-1472 + 2561 \gamma)) x^4 + 158203125 x^5$.	
	\begin{figure}[ht]
		\centering
		\includegraphics[width=0.7\linewidth]{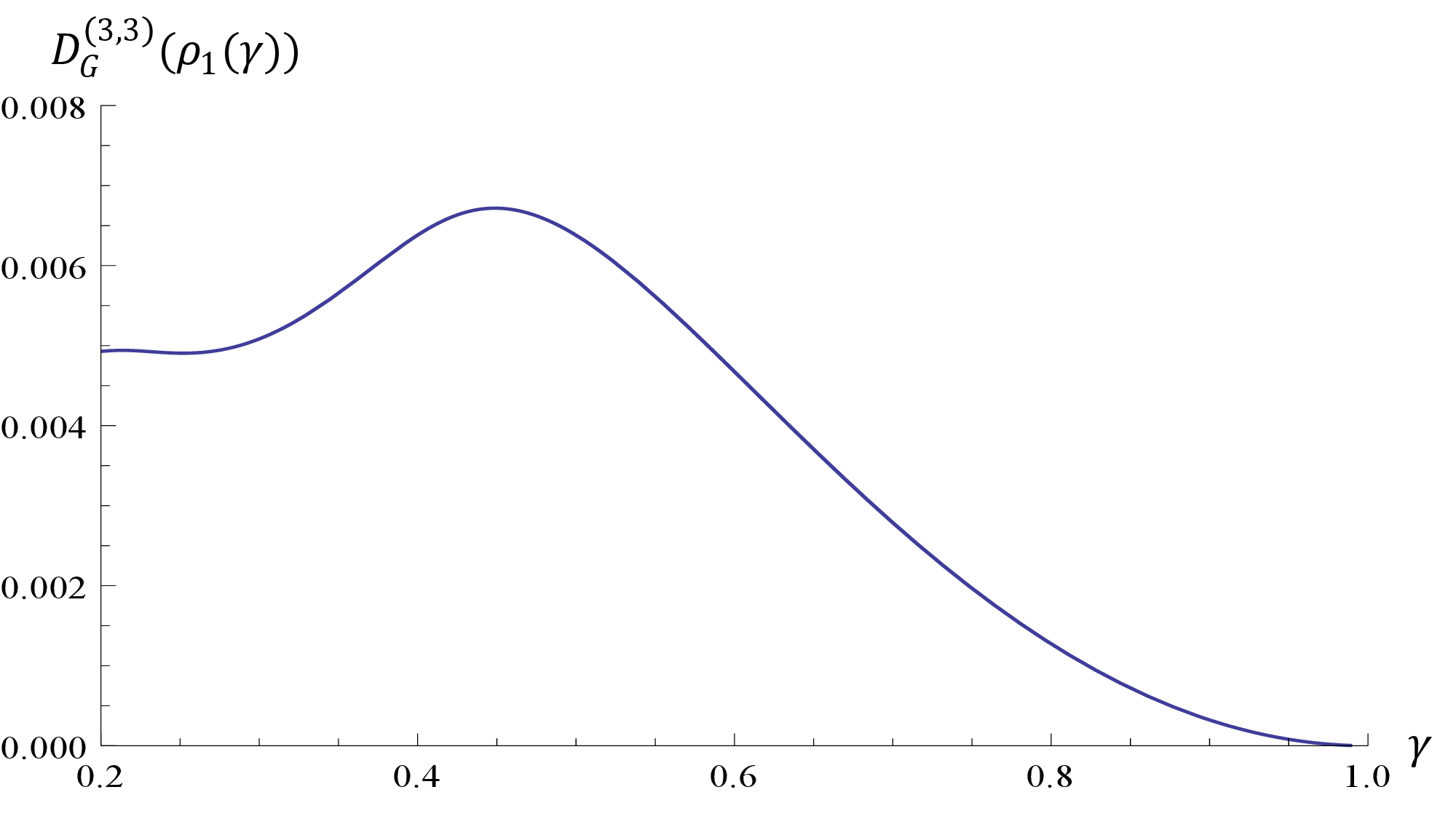}
		\caption{Geometric discord for the two-qutrit separable state $\rho_1(\gamma)$, defined in (\ref{sep_rhogamma}) is plotted. The $X$ axis represents the range of the state parameter $\gamma\in [\frac{1}{5},1]$, and the $Y$ axis represents the corresponding geometric discord given by (\ref{gd_sep_rhogammai}).}
		\label{fig:1our_sep_rhogammai}
	\end{figure}
	It can be observed from FIG. \ref{fig:1our_sep_rhogammai} that geometric discord may increase and decrease as the value of the state parameter increases from $\frac{1}{5}$ and approaches to 1.\\
	\subsubsection{Geometric discord of NPT Entangled States}
	Let us consider a few examples of NPTES, i.e., those states whose partial transposition matrix have at least one negative eigenvalue. We will now calculate the geometric discord of NPTES.\\
	\textbf{Example 1:} Let us call back the isotropic state $\rho_{\beta}$ 
	\begin{align}
		\begin{split}
			\rho_\beta=&\beta \ket{\phi^+}\bra{\phi^+}+\frac{1-\beta}{9} I_9, ~\frac{1}{3}\leq \beta \leq 1
		\end{split}
		\label{nptes_rho_beta}
	\end{align}
	where $I_9$ denotes the $9\times 9$ identity matrix and $\ket{\phi^+}$ represents the two-qutrit Bell-state, expressed as $\frac{1}{\sqrt{3}}(\ket{00}+\ket{11}+\ket{22})$. The isotropic state $\rho_{\beta}$ is a NPT entangled state when $\frac{1}{3}\leq \beta \leq 1$. The geometric discord of this particular state (\ref{nptes_rho_beta}) is given by 
	\begin{align}
		D_G^{(3,3)}(\rho_{\beta})=\frac{32}{243}\beta^2
		\label{gd_nptes_rho_isotropic}
	\end{align}
	
	\begin{figure}[ht]
		\centering
		\includegraphics[width=0.7\linewidth]{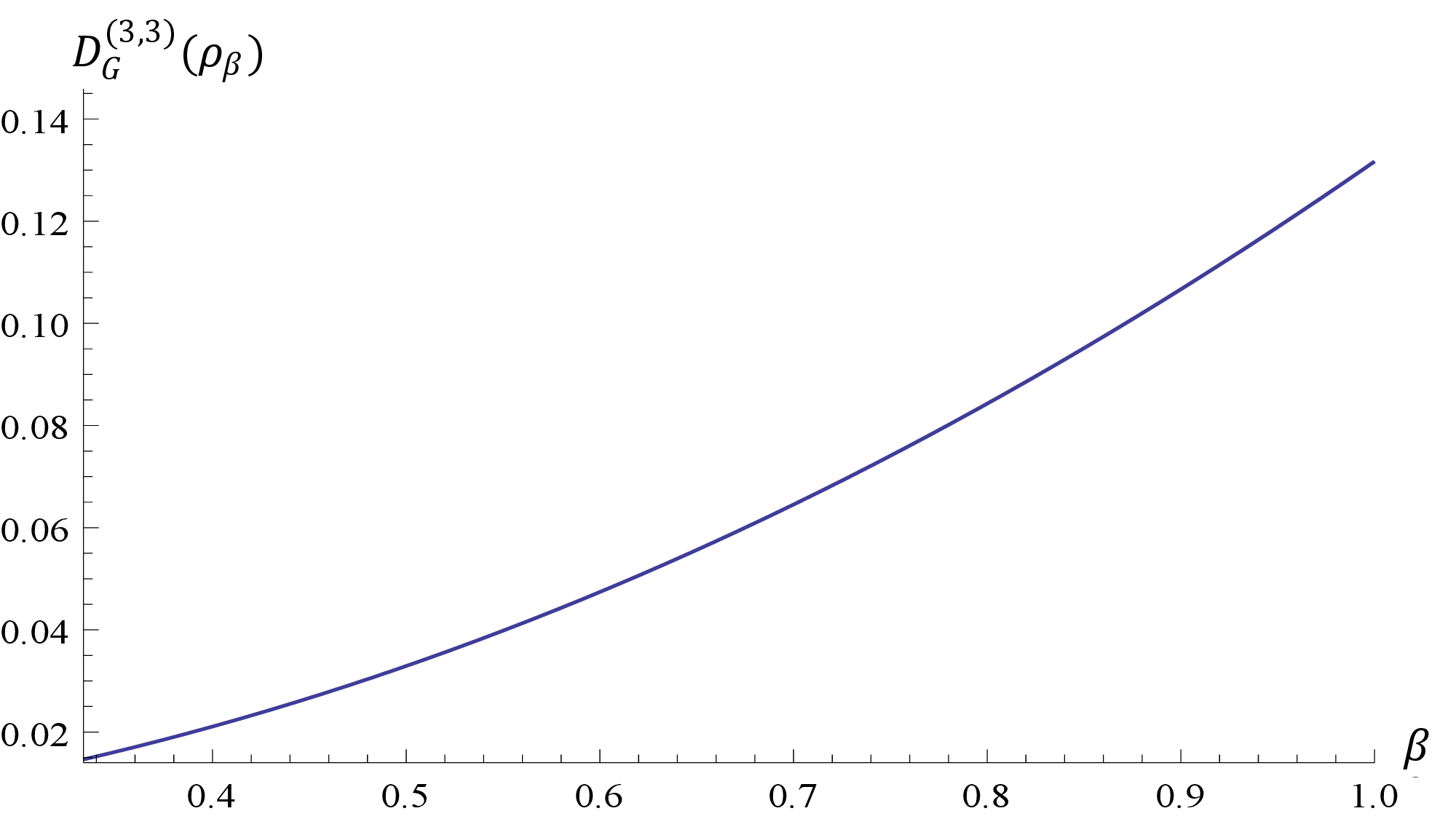}
		\caption{Geometric discord for the NPTES isotropic state $\rho_{\beta}$, where $\beta\in [\frac{1}{3},1]$ is plotted. The $X$ axis represents the state parameter $\beta$ lying in the interval $[\frac{1}{3},1]$, and the $Y$ axis represents the corresponding geometric discord $D_G^{(3,3)}(\rho_{\beta})$ given by (\ref{gd_nptes_rho_isotropic}).}
		\label{fig:1our_nptes_isotropic}
	\end{figure}
	In this case, we can see from FIG. \ref{fig:1our_nptes_isotropic} that the geometric discord is an increasing function of the state parameter $\beta$.\\
	\textbf{Example 2:} Let us recall the class of bipartite quantum state $\rho_{\alpha}$, given in (\ref{alpha_state}). The state $\rho_{\alpha}$ may now be defined for different range of $\alpha$, which is given as
	\begin{align}
		\rho_{\alpha}=&\frac{2}{7}\ket{\psi^+}\bra{\psi^+}+\frac{\alpha}{7}\sigma^++\frac{5-\alpha}{7}\sigma^-,~ 4< \alpha\leq 5
		\label{4alpha}
	\end{align}
	The state $\rho_{\alpha}$ is NPTES when $\alpha\in(4,5]$. 
	The geometric discord of the state $\rho_{\alpha}$ is given by (\ref{our_NPTES_alpha_d_proof}).
	\begin{align}
		D^{(3,3)}_G(\rho_{\alpha})=\frac{128}{11907},~ 4<\alpha\leq 5
		\label{gd_nptes_rho_alpha}
	\end{align}
	Here, it is interesting to observe here the fact that the geometric discord of the state $\rho_{\alpha}$ is independent of the parameter $\alpha$, when $\alpha\in (4,5]$.\\ 
	\textbf{Example 3:}  Now, let us examine another two-qutrit state, described by the density operator 
	\begin{align}
		\begin{split}
			\rho_a=&\frac{1}{5+2a^2}\sum_{i=1}^{3}\ket{\chi_i}\bra{\chi_i},~\frac{1}{\sqrt{2}}\leq a\leq 1
		\end{split}
		\label{nptes_rhoa}
	\end{align}
	where $\ket{\chi_i}=\ket{0i}-a\ket{i0},~i=1,2$ and $\ket{\chi_3}=\sum_{j=0}^{2}\ket{jj}$ denote the unnormalized state. Therefore, the geometric discord of the state (\ref{nptes_rhoa}) is given by
	\begin{align}
		D^{(3,3)}_G(\rho_a)=&\frac{8(128+a(17a^3-8a-72))}{729(5+2a^2)^2},~\frac{1}{\sqrt{2}}\leq a\leq 1
		\label{gd_nptes_rhoa}
	\end{align}
	
	\begin{figure}[ht]
		\centering
		\includegraphics[width=0.7\linewidth]{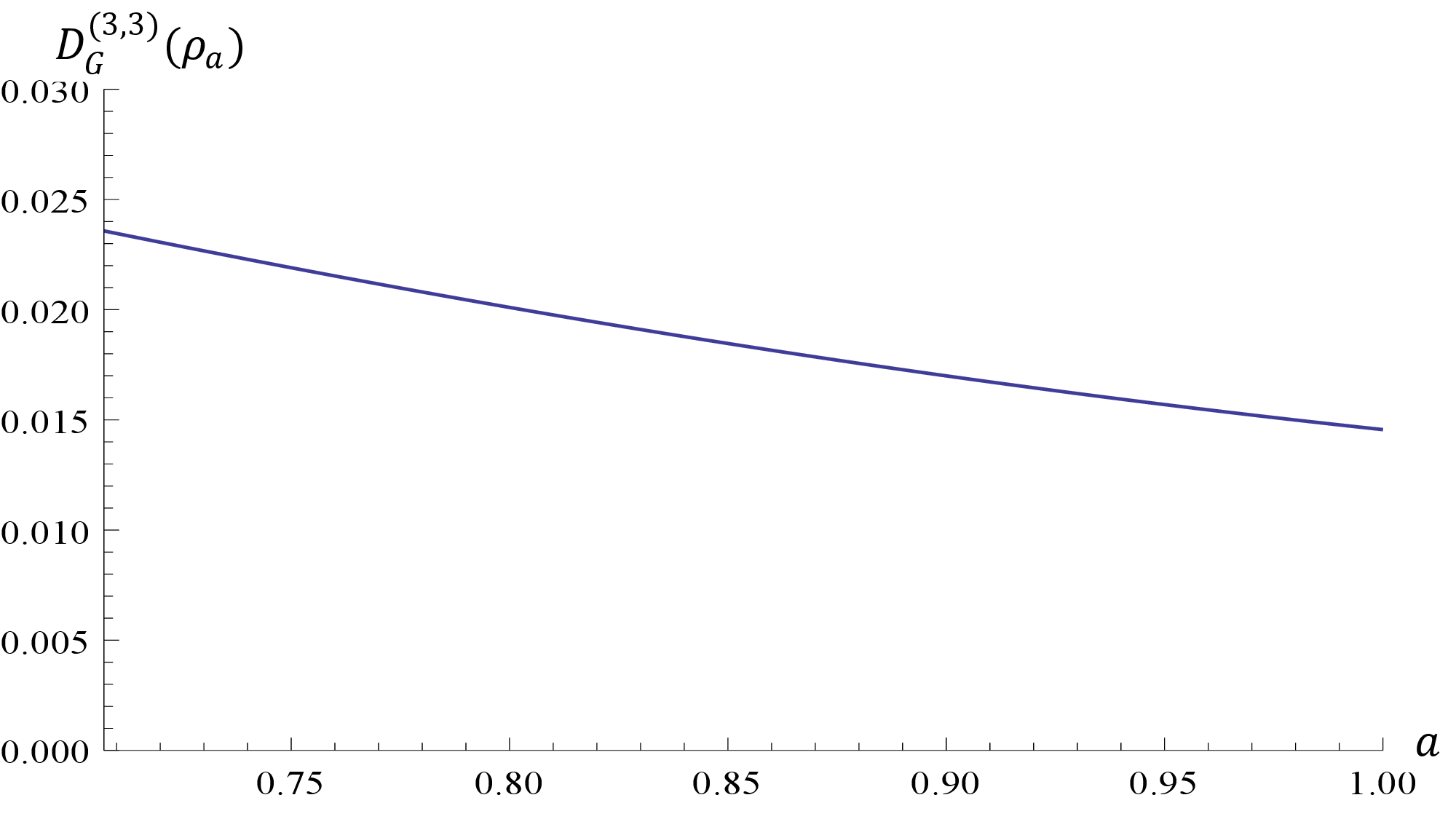}
		\caption{Geometric discord of the state $\rho_a$ in the interval $\frac{1}{\sqrt{2}}\leq a\leq 1$, is plotted. The $X$ axis represents the state parameter $a$, and the $Y$ axis represents the corresponding geometric discord $D_G^{(3,3)}(\rho_a)$ given by (\ref{gd_nptes_rhoa}).}
		\label{fig:1our_nptes_rhoa}
	\end{figure}
	From FIG. \ref{fig:1our_nptes_rhoa}, we can say that the quantum discord decreases in the interval $a \in [\frac{1}{\sqrt{2}},1]$, where the state under investigation is defined.\\
	\textbf{Example 4:} Consider a $4\otimes 4$ bipartite quantum state $\rho^{4\otimes4}_{cons}$ which is given by
	\begin{align}
		\rho_{cons}^{4\otimes 4}=&\frac{1}{4}\left(\sum_{i,j=0}^{3}\ket{ii}\bra{jj}\right)
		\label{rho4t4}
	\end{align}
	The geometric discord of the NPT entangled state $\rho_{cons}^{4\otimes 4}$ is given by
	\begin{align}
		D_G^{(4,4)}(\rho_{cons}^{4\otimes 4})=\frac{3}{64}
	\end{align}
	
	In the above discussed examples of various NPTES, we observe that the nature of geometric discord varies from state to state. It may be an increasing, decreasing or a constant function of the given state parameter. \\
	\subsubsection{Geometric discord of PPT Entangled States}
	In this subsection, we will consider some entangled states whose partial transposition matrix does not have any negative eigenvalues, i.e., all eigenvalues of partial transpose matrix are positive. These states are known as the positive partial transpose entangled states (PPTES). Now, our task is to calculate the geometric discord of a few PPT entangled states.\\
	\textbf{Example 1:} Let us consider the density operator $\rho_c$, given by \cite{horodecki_1997}
	\begin{align}
		\rho_c=\frac{1}{8c+1}\begin{pmatrix}
			c & 0 & 0 & 0 & c & 0 & 0 & 0 & c \\
			0 & c & 0 & 0 & 0 & 0 & 0 & 0 & 0 \\
			0 & 0 & c & 0 & 0 & 0 & 0 & 0 & 0 \\
			0 & 0 & 0 & c & 0 & 0 & 0 & 0 & 0 \\
			c & 0 & 0 & 0 & c & 0 & 0 & 0 & c \\
			0 & 0 & 0 & 0 & 0 & c & 0 & 0 & 0 \\
			0 & 0 & 0 & 0 & 0 & 0 & \frac{1+c}{2} & 0 & \frac{\sqrt{1+c^2}}{2} \\
			0 & 0 & 0 & 0 & 0 & 0 & 0 & c & 0 \\
			c & 0 & 0 & 0 & c & 0 & \frac{\sqrt{1+c^2}}{2} & 0 & \frac{1+c}{2}
		\end{pmatrix}
	\end{align}
	The state $\rho_{c}$ is PPTES in the range $0< c< 1$. The geometric discord of $\rho_{c}$ is given by
	\begin{align}
		D_G^{(3,3)}(\rho_c)=\begin{cases}
			\frac{8(125c^2-22c+17)}{729(1+8c)^2}-x_1-x_2& 0 < c<0.4142\\
			\frac{8(107c^2-22c+17)}{729(1+c)^2}-x_1 &0.4142<c < 1
		\end{cases}
		\label{gd_pptes_rhoc}
	\end{align}
	where, $x_1>x_2>x_3$ are the three roots of the cubic equation $x^3+(-34+44c-70c^2)x^2+(270-648c+1548c^2-2232c^3+1062c^4)x+(-1944c^2+7776c^3-11664c^4+7776c^5-1944c^6)=0$.\\	
	\begin{figure}[ht]
		\centering
		\includegraphics[width=0.7\linewidth]{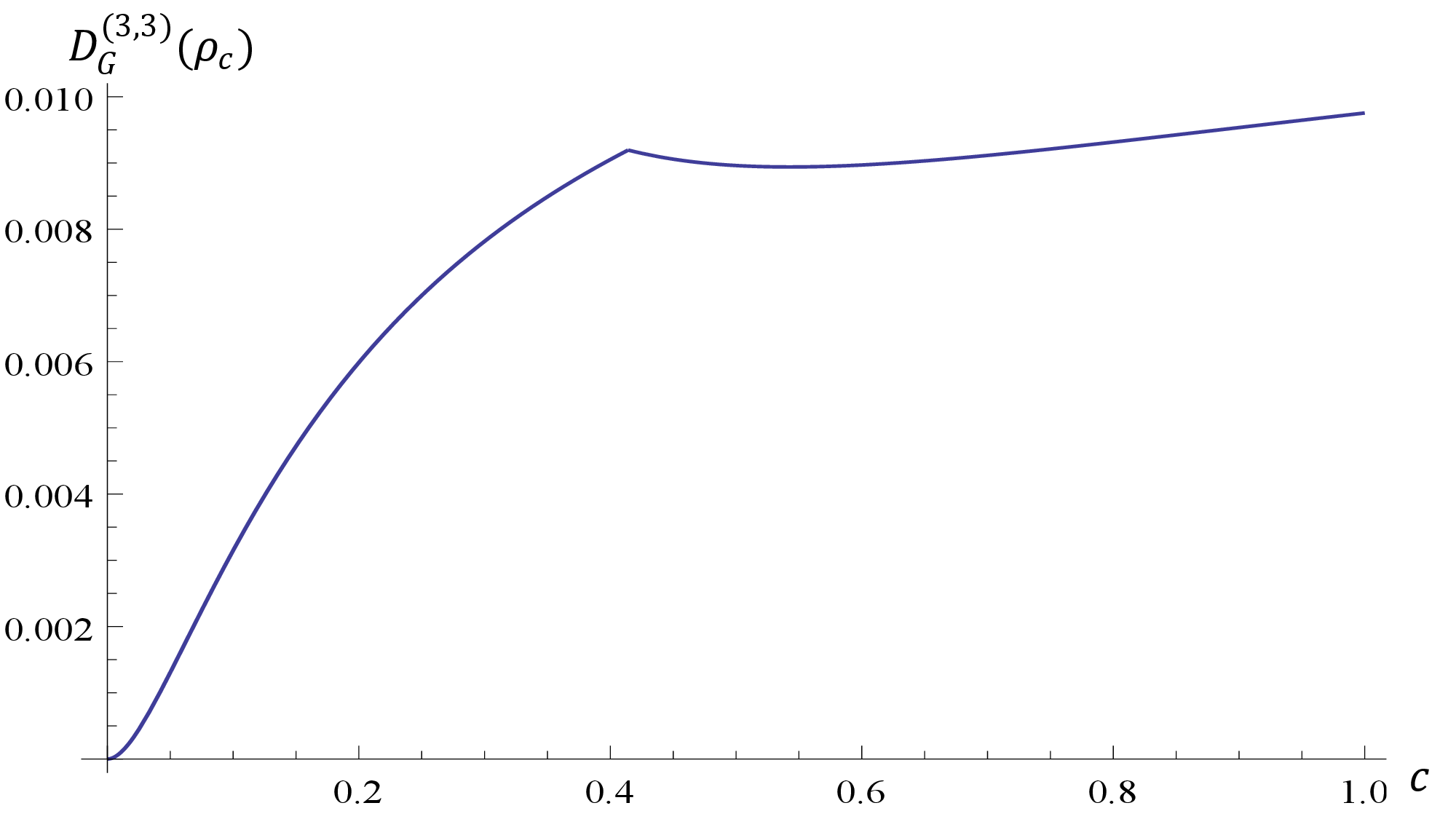}
		\caption{Geometric discord for the state $\rho_c$, where $0<c<1$ is plotted. The $X$ axis represents the state parameter $c$ lying in the interval $(0,1)$, and the $Y$ axis represents the corresponding geometric discord $D_G^{(3,3)}(\rho_c)$ given by (\ref{gd_pptes_rhoc}).}
		\label{fig:1our_pptes_rhoc}
	\end{figure}
	
	It can be observed from FIG. \ref{fig:1our_pptes_rhoc} that the curve representing geometric discord $D^{(3,3)}_G(\rho_{c})$ increases when the state parameter $c\in(0,0.4142)$, but varies in the range $c\in(0.4142,1)$.\\

	\textbf{Example 2:} Now, again recalling the state $\rho_{\alpha}$ defined in (\ref{alpha_state}) as
	\begin{align}
		\rho_{\alpha}=&\frac{2}{7}\ket{\psi^+}\bra{\psi^+}+\frac{\alpha}{7}\sigma^++\frac{5-\alpha}{7}\sigma^-,~ 3< \alpha\leq 4
	\end{align}
	
	The state $\rho_{\alpha}$ is a PPTES when $\alpha\in (3,4]$. The geometric discord of $\rho_{\alpha}$ in this case is given by (Refer to Appendix D for detailed calculations (\ref{our_PPTES_alpha_d_proof}))
	\begin{align}
		D^{(3,3)}_G(\rho_{\alpha})=
		\begin{cases}
			\frac{32}{11907}(\alpha^2-5\alpha+9)& 3< \alpha \leq 3.618\\
			\frac{128}{11907} & 3.618<\alpha\leq 4
		\end{cases}
		\label{gd_pptes_rho_alpha}
	\end{align}
	
	\begin{figure}[ht]
		\centering
		\includegraphics[width=0.7\linewidth]{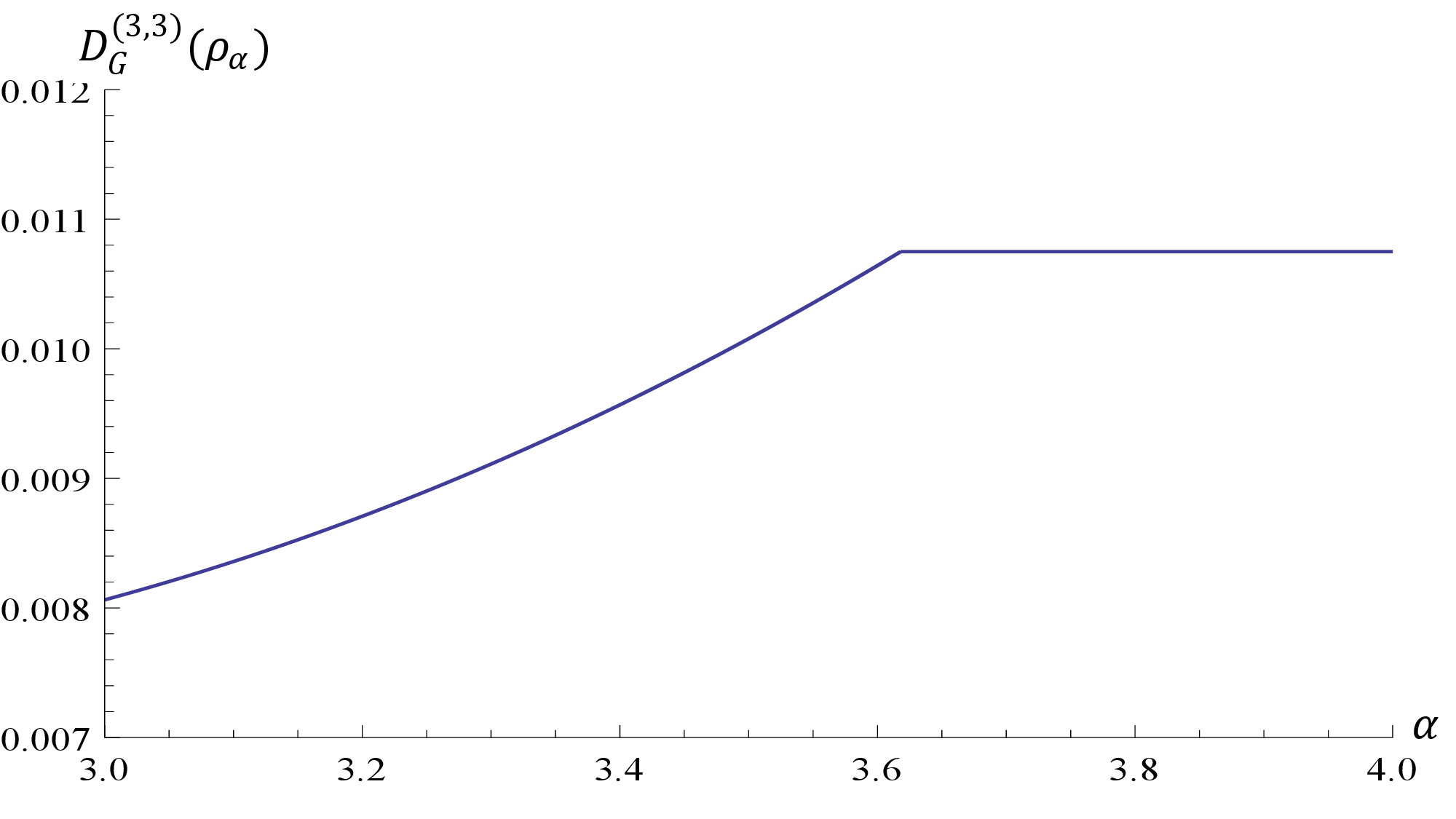}
		\caption{Geometric discord for the state $\rho_{\alpha}$, where $3< \alpha \leq 4$ is plotted. The $X$ axis represents the value of the state parameter $\alpha$ lying in the interval $(3,4]$, and the $Y$ axis represents the corresponding geometric discord $D_G^{(3,3)}(\rho_{\alpha})$ given by (\ref{gd_pptes_rho_alpha}).}
		\label{fig:1our_pptes_alpha}
	\end{figure}
	From FIG. \ref{fig:1our_pptes_alpha}, we find that the geometric discord of $\rho_{\alpha}$ when $\alpha\in (3,3.618]$, increases and it becomes constant when $\alpha\in(3.618,4]$.\\

	\textbf{Example 3:} Let us take the following PPT entangled state, which can be described by the density operator $\rho_{cons}^{3\otimes 3}$ as 
	\begin{align}
		\rho_{cons}^{3\otimes 3}=&\begin{pmatrix}
			a & 0 & 0 & 0 & b & 0 & 0 & 0 & b \\
			0 & c & 0 & 0 & 0 & 0 & 0 & 0 & 0 \\
			0 & 0 & a & 0 & 0 & 0 & 0 & 0 & 0 \\
			0 & 0 & 0 & a & 0 & 0 & 0 & 0 & 0 \\
			b & 0 & 0 & 0 & a & 0 & 0 & 0 & 0 \\
			0 & 0 & 0 & 0 & 0 & c & 0 & b & 0 \\
			0 & 0 & 0 & 0 & 0 & 0 & c & 0 & 0 \\
			0 & 0 & 0 & 0 & 0 & b & 0 & a & 0 \\
			b & 0 & 0 & 0 & 0 & 0 & 0 & 0 & a
		\end{pmatrix}
	\end{align}
	where $a=\frac{1+\sqrt{5}}{3+9\sqrt{5}}$, $b=\frac{-2}{3+9\sqrt{5}}$ and $c=\frac{-1+\sqrt{5}}{3+9\sqrt{5}}$. The geometric discord of $\rho_{cons}^{3\otimes 3}$ is given by
	\begin{align}
		D_G^{(3,3)}(\rho_{cons}^{3\otimes 3})=&\frac{16(23-3\sqrt{5})}{29403}>0
	\end{align}

	\section{Relationship between geometric discord and negativity}
	In this section, we will consider a few example to find out whether there exist any relationship between negativity, which is a measure of entanglement and the geometric discord, which is a measure of general quantum correlation. We have to proceed with NPTES as negativity vanishes for all PPTES. Therefore, let us first define the negativity for a $d\otimes d$-dimensional NPTES system described by the density matrix $\rho^{NPTES}$. It is defined as 
	\begin{align}
		N(\rho^{NPTES})&=\frac{||(\rho^{NPTES})^{T_B}||_1-1}{d-1}\nonumber\\&=\frac{2}{d-1}\sum_{\lambda_i<0}|\lambda_i((\rho^{NPTES})^{T_B})|
		\label{negativity}
	\end{align}
	where $||.||_1$ denotes the trace norm and $(\rho^{NPTES})^{T_B}$ is the matrix obtained by taking the partial transposition of $\rho^{NPTES}$. It can be noted here that we can use other measures of entanglement as well, such as concurrence. But in higher dimensional mixed systems, concurrence has no closed formula. On the other hand, unlike concurrence, negativity can be calculated using a simple formula, which works for any arbitrary dimensional bipartite system.\\
	
	\textbf{Example 1:} Let us recall $\alpha$ state, defined in (\ref{4alpha}). The state is NPTES in the interval $\alpha\in(4,5]$ and it may be described by the density operator $\rho_{\alpha}^{NPTES}$. Therefore, to quantify the amount of entanglement in it, we first find the negative eigenvalues of partial transpose of $\rho_{\alpha}^{NPTES}$ in the range $\alpha\in (4,5]$ and then we use the formula (\ref{negativity}) to calculate the negativity. The negative eigenvalue is $\frac{1}{42}(5-\sqrt{41-20a+4a^2})$ with multiplicity 3. Therefore, the negativity of the state $\rho_{\alpha}^{NPTES}$ is given by
	\begin{align}
		N(\rho_{\alpha}^{NPTES})=\frac{1}{14}(5-\sqrt{41-20a+4a^2})
	\end{align}
	\begin{figure}[ht]
		\centering
		\includegraphics[width=0.7\linewidth]{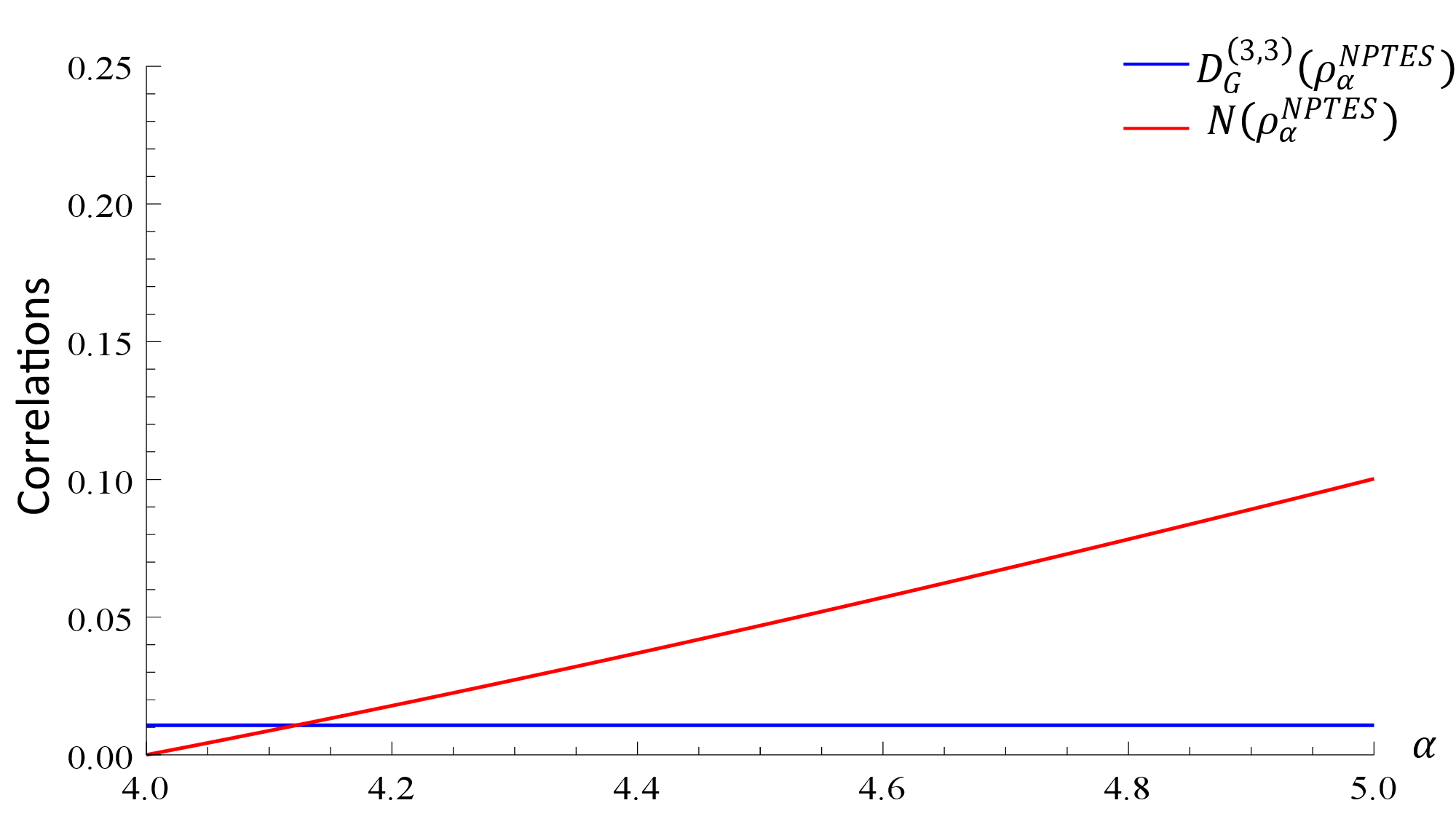}
		\caption{Graph depicting the curves for geometric discord and negativity for $\rho_{\alpha}^{NPTES}$ when $\alpha\in (4,5]$. In this plot, X-axis represents the state parameter $\alpha$.}
		\label{our_nptes_alpha}
	\end{figure}
	The geometric discord of the state $\rho_{\alpha}^{NPTES}$ is $D^{(3,3)}_G(\rho_{\alpha}^{NPTES})=\frac{128}{11907},~ 4<\alpha\leq 5$, which is given in (\ref{gd_nptes_rho_alpha}).\\ 
	It can be clearly observed from Fig. (\ref{our_nptes_alpha}) that in the range $\alpha\in (4,4.12)$, quantum discord is greater than negativity. But in the range, $\alpha\in (4.12,5]$, negativity is greater than quantum discord. At the point $\alpha=4.12$, geometric discord and the negativity coincide. \\
	
	\textbf{Example 2:} Consider the isotropic state defined in (\ref{nptes_rho_beta}) and it is denoted by $\rho_{\beta}^{NPTES}$ in this example. It is indeed a NPTES. Therefore, the geometric discord of $\rho_{\beta}^{NPTES}$ can be calculated as
	\begin{align}
		D_G^{(3,3)}(\rho_{\beta}^{NPTES})=\frac{32}{243}\beta^2
	\end{align}
	Since the state $\rho_{\beta}$ is a NPTES, we can quantify its entanglement through negativity. The negative eigenvalues of partial transpose of $\rho_{\beta}$ is $\frac{1-4\beta}{9}$ with multiplicity 2. Therefore, we have 
	\begin{align}
		N(\rho_{\beta}^{NPTES})=\frac{2}{9}(1-4\beta)
	\end{align}
	
	\begin{figure}[ht]
		\centering
		\includegraphics[width=0.7\linewidth]{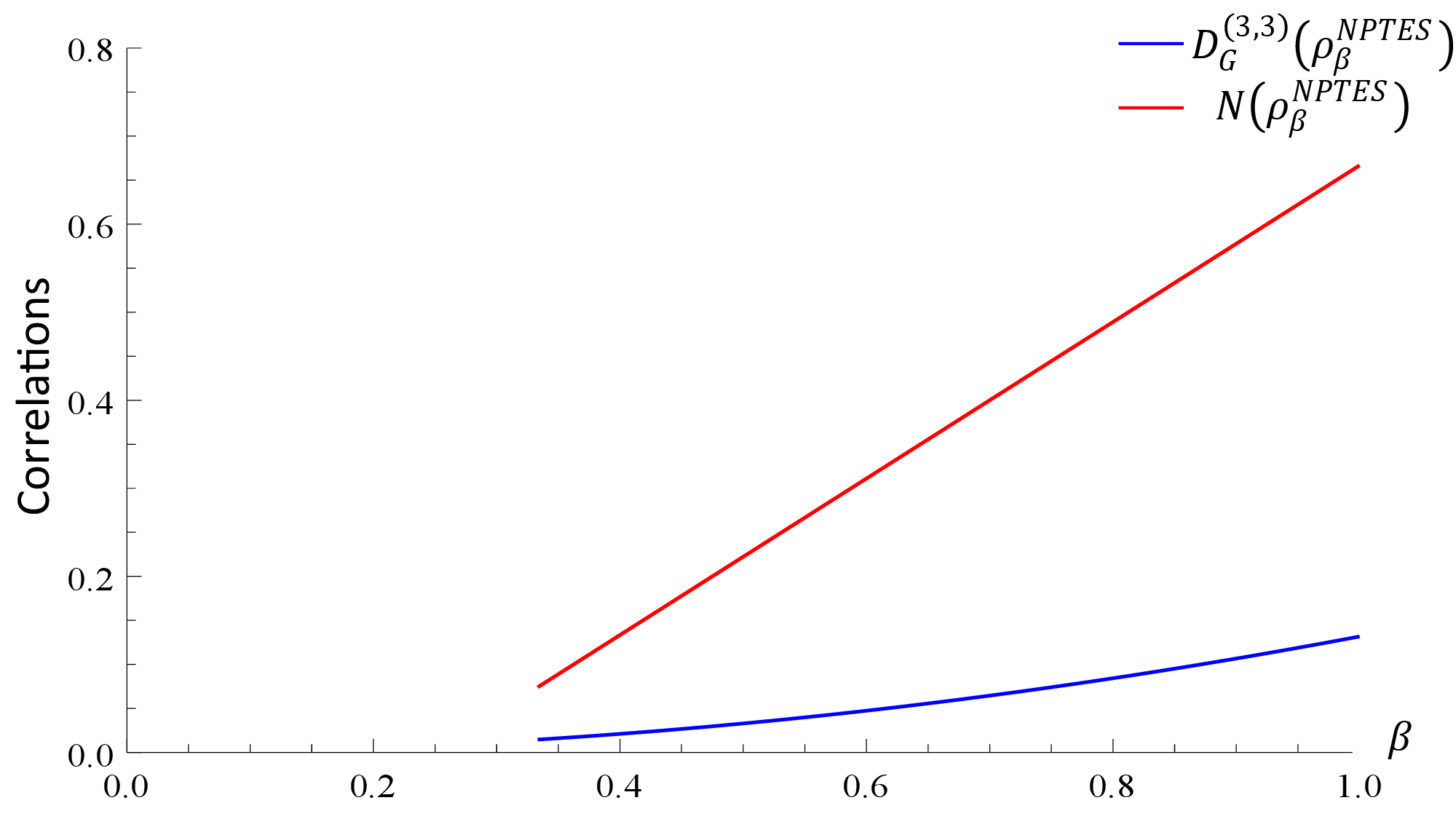}
		\caption{Graph depicts the comparison between geometric discord and negativity for $\rho_{\beta}$. Here, X-axis represents the state parameter $\beta$.}
		\label{our_nptes_isotropic}
	\end{figure}
	In this case, it can be observed that the negativity and geometric discord never coincide, unlike in Example-1 of this section. Moreover, in this case, the negativity is always greater than the geometric discord. \\
	\textbf{Example 3:} Let us consider the state defined in (\ref{nptes_rhoa}) and it may be renamed as $\rho_a^{NPTES}$. The geometric discord of the state $\rho_a^{NPTES}$ is given by
	\begin{align}
		D^{(3,3)}_G(\rho_a^{NPTES})=&\frac{8(128+a(17a^3-8a-72))}{729(5+2a^2)^2},~\frac{1}{\sqrt{2}}\leq a\leq 1
	\end{align}
	Further, we find that there are two negative eigenvalues of partial transpose of $\rho_a^{NPTES}$ and they are listed as $\frac{1-\sqrt{2}a}{5+2a^2}$ and $\frac{1+a^2-\sqrt{5-2a^2+a^4}}{2(5+2a^2)}$, each with multiplicity 1. Therefore, we get
	\begin{align}
		N(\rho_a^{NPTES})=\frac{1-\sqrt{2}a}{5+2a^2}+\frac{1+a^2-\sqrt{5-2a^2+a^4}}{2(5+2a^2)}
	\end{align}

	\begin{figure}[ht]
		\centering
		\includegraphics[width=0.7\linewidth]{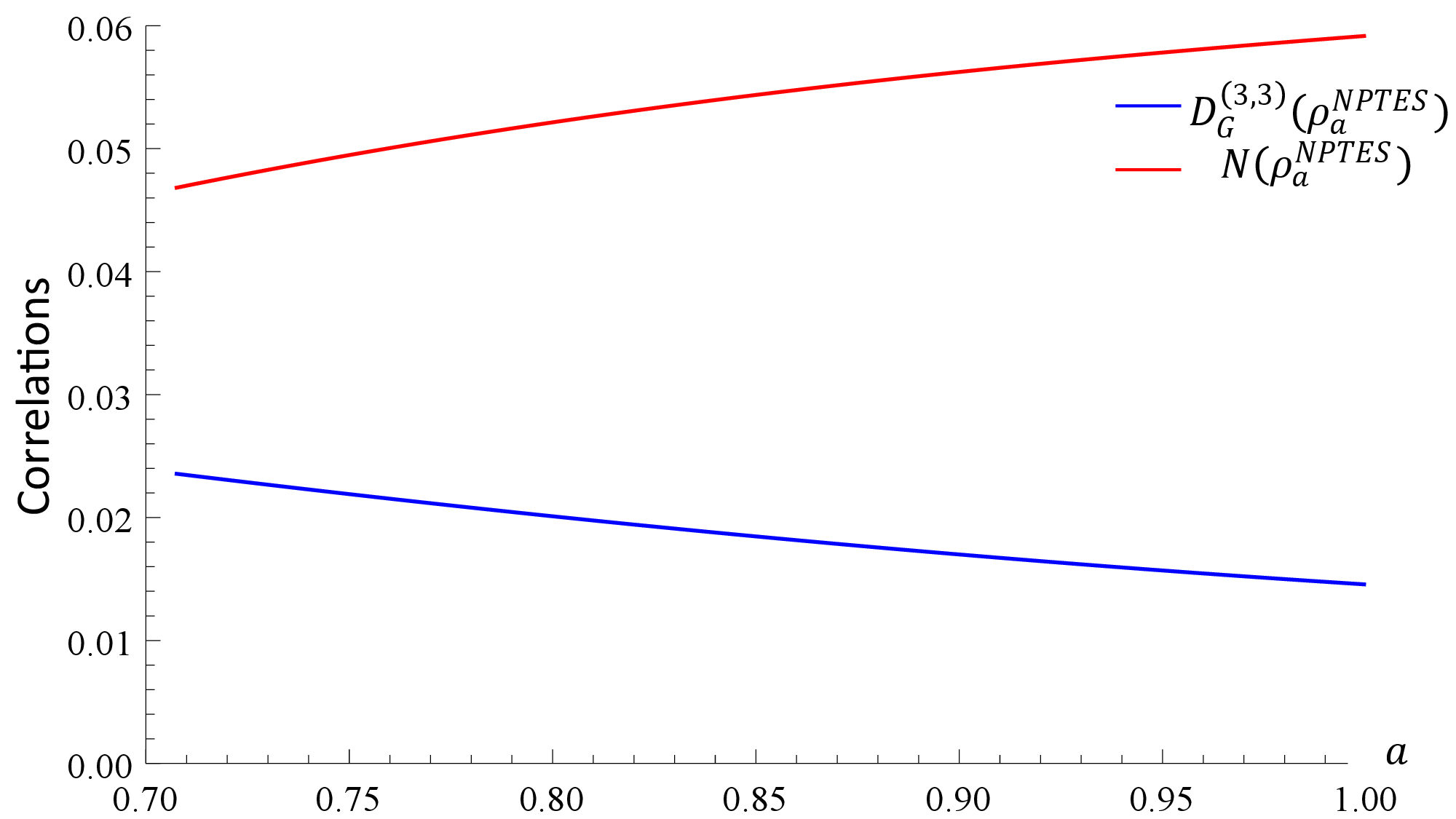}
		\caption{Geometric discord and negativity has been studied for $\rho_a^{NPTES}$. The value of the parameter $a$ goes along the X-axis.}
		\label{our_nptes_rhoa}
	\end{figure}
	From Fig. (\ref{our_nptes_rhoa}), it can be seen that negativity increases, whereas the geometric discord decreases as the state parameter $\alpha$ increases from $\frac{1}{\sqrt{2}}$ to $1$. Thus, we can conclude for this example that negativity is always greater than the geometric discord.
	
	\section{Applications}
	In a QKD protocol, two distant parties establish a secret key using quantum channel in presence of an eavesdropper. In this section, we will focus on an important application of GQD in context of key distribution protocols. In particular, we will discuss the case where the communicating parties make use of private states for generation of the secret key, irrespective of whether the entanglement is present or absent between the sender and the receiver.
	\subsection{Security in quantum states}
	It is well known that the two distant parties, namely Alice and Bob can share a secret using QKD scheme when they share a singlet state \cite{ekert_1991}. In this case, Alice and Bob perform local operations on the part of the singlet state in their possession and obtain a secret random sequence of bits, known as the key. This key is completely uncorrelated with any noise that maybe present in the environment. The key distribution is also possible when the shared state between Alice and Bob is a mixed state and the key is hidden in a part of the shared state \cite{Horodecki_2008}.\\
	Consider the four-qubit mixed state $\rho_{ABA'B'}\in B(\mathbb{C}^d\otimes \mathbb{C}^d \otimes \mathbb{C}^d \otimes \mathbb{C}^d)$ shared between Alice and Bob, where the qubits $A$ and $A'$ belong to Alice and the qubits $B$ and $B'$ are possessed by Bob. The four-qubit mixed state $\rho_{ABA'B'}$ with respect to a basis $B_1\equiv\{\ket{e_if_j}_{AB}\}_{i,j=1}^{d-1}$ can be expressed in the form \cite{Horodecki_2008}
	\begin{align}
		\rho_{ABA'B'}=\sum_{i,j,k,l=0}^{d-1}\ket{e_if_j}_{AB}\bra{e_kf_l} \otimes A_{{ijkl}_{A'B'}}
		\label{rho ABA'B'}
	\end{align}
	The subsystem $AB$ of the complete system described by the density operator $\rho_{ABA'B'}$ is known as the \textit{key part}, from which the secret key can be obtained after von Neumann measurements are performed on the local subsystems. The subsystem $A'B'$ is known as the \textit{shield} of the state, which provides the security to the key part from an eavesdropper. In order to extract the key from the shared state $\rho_{ABA'B'}$, Alice and Bob perform local measurements on the subsystem $AB$ and discuss their outcomes through the public channel.\\
	The classical-classical-quantum (ccq) state associated with the state $\rho_{ABA'B'}$ is given by \cite{Horodecki_2008}
	\begin{align}
		\rho_{ABE}^{ccq(ideal)}=\sum_{i,j=0}^{d-1}\frac{1}{d}\ket{e_if_j}_{AB}\bra{e_if_j}\otimes \rho_E
	\end{align}
	where $\rho_E$ denote the density operator possessed by the eavesdropper, namely Eve.\\
	Now, consider an unitary operator $U$ of the form
	\begin{align}
		U=\sum_{k,l=0}^{d-1}\ket{e_kf_l}_{AB}\bra{e_kf_l}\otimes U^{kl}_{A'B'}
	\end{align}
	This operator is known as \textit{twisting} operator which is controlled by the basis $B_1$ \cite{Horodecki_2005_prl_94}.\\
	For every state $\rho_{ABA'B'}$, there exists a twisting operator $U$ such that upon applying it to the given state, and tracing out $A'B'$, we obtain $\rho_{AB}=\text{Tr}_{A'B'}[U\rho_{ABA'B'}U^\dagger]$. This operation is called privacy-squeezing, and the output state $\rho_{AB}$ is known as the privacy-squeezed state.\\
	The state obtained after applying the privacy-squeezing is given by \cite{Horodecki_2018}
	\begin{align}
		\sigma_{ABA'B'}=\frac{1}{d} \sum_{i,j=0}^{d-1}\ket{e_if_j}_{AB}\bra{e_if_j}\otimes U_i \sigma_{A'B'}U_j^\dagger
		\label{ps_sigma_aba'b'}
	\end{align}
	This state is known as the \textit{private state}. These states possesses the property that the secret key can be accessed by performing measurement on the key part, i.e., $AB$ subsystem in the computational basis. The ccq state obtained after privacy squeezing operation is such that Eve's system is in product state with the key part. Therefore, the ccq state of (\ref{ps_sigma_aba'b'}) obtained after privacy squeezing is given by
	\begin{align}
		\sigma_{ABE}^{ccq(squeezing)}=\sum_{i,j=0}^{d-1}p_{ij}\ket{e_if_j}_{AB}\bra{e_if_j}\otimes \rho_{E}
	\end{align}
	
	The benefit of employing private states in a QKD protocol is that the secret key can also be extracted from the bound entangled states, from which no singlets can be distilled. This explains the fact that non-distillable entangled states can have a great cryptographic value. The quantitative relation between secrecy and entanglement can be given using the private states \cite{Horodecki_2018}. In this work, we aim to extend this approach to a broader class of quantum correlations, specifically GQD. In the following subsection, we establish a connection between the distillable key and the geometric discord.

	\subsection{Relationship between distillable key rate $K_D$ and geometric discord $D^{(3,3)}_G(\rho)$}
	
	Let the four-qubit state $\rho_{ABA'B'}\in B(\mathbb{C}^2\otimes \mathbb{C}^2 \otimes \mathbb{C}^d \otimes \mathbb{C}^d)$ shared between Alice and Bob is of the form 
	\begin{align}
		\begin{split}
			\rho_{ABA'B'}=&\ket{\phi^+}_{AB}\bra{\phi^+}\otimes (\sigma_0)_{A'B'} +\ket{\phi^-}_{AB}\bra{\phi^-}\otimes (\sigma_1)_{A'B'} +\\
			&\ket{\psi^+}_{AB}\bra{\psi^+}\otimes (\sigma_2)_{A'B'} +\ket{\psi^-}_{AB}\bra{\psi^-}\otimes (\sigma_3)_{A'B'}
		\end{split}
		\label{shared state kd}
	\end{align}
	where $\ket{\phi^\pm}=\frac{1}{\sqrt{2}}(\ket{00}\pm \ket{11})$ and $\ket{\psi^\pm}=\frac{1}{\sqrt{2}}(\ket{01}\pm \ket{10})$ are the Bell states in $\mathbb{C}^2\otimes \mathbb{C}^2$. Further, $\sigma_i$'s $(i=0,1,2,3)$ denote the $d^2$-dimensional density matrices. The qubits $A$ and $A'$ belongs to Alice and the qubits $B$ and $B'$ belongs to Bob. The subsystem $AB$ contains the key part, which is protected by the shield provided by subsystem $A'B'$.\\
	The state (\ref{shared state kd}) can be re-expressed in the matrix form as
	\begin{align}
		\rho_{ABA'B'}=\frac{1}{2} \begin{bmatrix}
			\sigma_0+\sigma_1 & 0 & 0 & \sigma_0-\sigma_1  \\
			0 & \sigma_2+\sigma_3  & \sigma_2-\sigma_3  & 0 \\
			0 & \sigma_2-\sigma_3  & \sigma_2+\sigma_3  & 0 \\
			\sigma_0-\sigma_1 	& 0 & 0 & \sigma_0+\sigma_1 
		\end{bmatrix}
	\end{align}
	Let us now apply twisting operation on the subsystem $A'B'$ and then after tracing out the subsystem $A'B'$, the privacy squeezed (p-squeezed) state $\sigma_{AB}$ of the state $\rho_{ABA'B'}$ is given by
	\begin{align}
		\sigma_{AB}=\frac{1}{2} \begin{bmatrix}
			\left|\left|\sigma_0+\sigma_1\right|\right|_1 & 0 & 0 & \left|\left|\sigma_0-\sigma_1\right|\right|_1  \\
			0 & \left|\left|\sigma_2+\sigma_3\right|\right|_1  & \left|\left|\sigma_2-\sigma_3\right|\right|_1  & 0 \\
			0 & \left|\left|\sigma_2-\sigma_3\right|\right|_1  & \left|\left|\sigma_2+\sigma_3\right|\right|_1  & 0 \\
			\left|\left|\sigma_0-\sigma_1\right|\right|_1	& 0 & 0 & \left|\left|\sigma_0+\sigma_1\right|\right|_1
		\end{bmatrix}
	\end{align} 	
	where $\left|\left|.\right|\right|_1$ denotes the trace norm.\\
	In order to check the security of the privacy squeezed state described by the density operator $\sigma_{AB}$, let us consider the purification of $\sigma_{AB}$. We can add the ancillary state as eavesdropper's state, who tries to access the key part in such a way that the privacy squeezed state is purified and therefore, its purification is given by
	\begin{align}
		\begin{split}
			\ket{\psi}_{ABE}=&\sqrt{x}\ket{\phi^+}_{AB}\ket{e_0}_E+\sqrt{y}\ket{\phi^-}_{AB}\ket{e_1}_E+\sqrt{z}\ket{\psi^+}_{AB}\ket{e_2}_E+\sqrt{w}\ket{\psi^-}_{AB}\ket{e_3}_E
		\end{split}
		\label{purify_psiabe}
	\end{align}
	where $\{\ket{e_0},\ket{e_1},\ket{e_2},\ket{e_3}\}$ form an orthonormal basis for the subsystem $E$ possessed by the eavesdropper and
	\begin{align}
		\begin{split}
			x=&\frac{1}{2}(\left|\left|\sigma_0+\sigma_1\right|\right|_1+\left|\left|\sigma_0-\sigma_1\right|\right|_1)\\
			y=&\frac{1}{2}(\left|\left|\sigma_0+\sigma_1\right|\right|_1-\left|\left|\sigma_0-\sigma_1\right|\right|_1)\\
			z=&\frac{1}{2}(\left|\left|\sigma_2+\sigma_3\right|\right|_1+\left|\left|\sigma_2-\sigma_3\right|\right|_1)\\
			w=&\frac{1}{2}(\left|\left|\sigma_2+\sigma_3\right|\right|_1-\left|\left|\sigma_2-\sigma_3\right|\right|_1)
		\end{split}
		\label{xyzw_chi}
	\end{align}
	The purified state $\ket{\psi}_{ABE}$ can be re-expressed as a ccq-state as \cite{chi_2007}
	\begin{align}
		\sigma_{ABE}^{ccq}=\frac{1}{2}\sum_{i,j=0}^{1}\ket{ij}_{AB}\bra{ij}\otimes (\delta_{ij})_{E}
	\end{align}
	where the subsystem $AB$ can be expressed in terms of the computational basis $\{\ket{00},\ket{01},\ket{10},\ket{11}\}$ and  $(\delta_{ij})_{E},i,j=0,1$ denote the projectors of the eavesdropper's system. The projectors $(\delta_{00})_{E}, (\delta_{01})_{E}, (\delta_{10})_{E}$ and $(\delta_{11})_{E}$ are given by
	\begin{align}
		\begin{split}
			(\delta_{00})_{E}=\sqrt{x}\ket{e_0}_E+\sqrt{y}\ket{e_1}_E\\
			(\delta_{01})_{E}=\sqrt{x}\ket{e_0}_E-\sqrt{y}\ket{e_1}_E\\
			(\delta_{10})_{E}=\sqrt{z}\ket{e_2}_E+\sqrt{w}\ket{e_3}_E\\
			(\delta_{11})_{E}=\sqrt{z}\ket{e_2}_E-\sqrt{w}\ket{e_3}_E
		\end{split}
	\end{align}
	where $x,y,z,w$ are given in (\ref{xyzw_chi}).\\
	Now we are in a position to derive the relationship between the distillable secret key rate $K_D$ and the geometric discord $D_G^{(d_1,d_2)}(\rho)$. To start the derivation, we need to use the following mathematical result:\\
	\textbf{Result-1 \cite{Bhatia_1997_book, peng_2015}:} For any two linear operators $C$ and $D$, the relation between the trace norm $\normsq{C}_1$ and the Hilbert-Schmidt norm $\normsq{D}_2$ is given by 
	\begin{eqnarray}
		\normsq{C}_1\geq \normsq{D}_2
		\label{result1}
	\end{eqnarray}
	Choose $d\otimes d$ dimensional operators $\sigma_{0}$,$\sigma_{1}$,$\sigma_{2}$ and $\sigma_{3}$ in such a way that it satisfies the following:
	\begin{itemize}
		\item[O1.] $\sigma_{i}$'s are hermitian for $i=0,1,2,3$.
		\item[O2.] $\sigma_{i}$'s are positive operators.
		\item[O3.] $\text{Tr}(\sigma_{i})=1,~~i=0,1,2,3$.
		\item[O4.] $\left|\left|\sigma_0+\sigma_1\right|\right|_1=\left|\left|\sigma_0-\sigma_1\right|\right|_1$ and $\left|\left|\sigma_2+\sigma_3\right|\right|_1=\left|\left|\sigma_2-\sigma_3\right|\right|_1$.
		\label{operator_sigmai}
	\end{itemize}
	Therefore, using the result (\ref{result1}), the expression of $x,y,z,w$ given in (\ref{xyzw_chi}) reduces to
	\begin{align}
			x=& \frac{1}{2}(\left|\left|\sigma_0+\sigma_1\right|\right|_1+\left|\left|\sigma_0+\sigma_1\right|\right|_1)=\left|\left|\sigma_0+\sigma_1\right|\right|_1\notag\\
			\geq&		 \left|\left|\sigma_0+\sigma_1\right|\right|_2\\
			y=&\frac{1}{2}(\left|\left|\sigma_0+\sigma_1\right|\right|_1-\left|\left|\sigma_0+\sigma_1\right|\right|_1)=0\\
			z=& \frac{1}{2}(\left|\left|\sigma_2+\sigma_3\right|\right|_1+\left|\left|\sigma_2+\sigma_3\right|\right|_1)=\left|\left|\sigma_2+\sigma_3\right|\right|_1\notag\\
			\geq&\left|\left|\sigma_2+\sigma_3\right|\right|_2\\
			w=&\frac{1}{2}(\left|\left|\sigma_2+\sigma_3\right|\right|_1-\left|\left|\sigma_2+\sigma_3\right|\right|_1)=0
		\end{align}
		To express $x$ and $z$ in terms in geometric discord, we have
		\begin{align}
			\frac{x}{2}\geq& \frac{\left|\left|\sigma_0+\sigma_1\right|\right|_2}{2}\\
			\frac{z}{2}\geq& \frac{\left|\left|\sigma_2+\sigma_3\right|\right|_2}{2}\\
	\end{align}
	Now, if there exists some $\sigma_{cl}\in C$, where $C$ is the set of zero-discord states, for which $\left|\left|\frac{\sigma_0+\sigma_1}{2}\right|\right|_2\geq \left|\left|\frac{\sigma_0+\sigma_1}{2}-\sigma_{cl}\right|\right|_2$, then we have
	\begin{align}
		\begin{split}
			\frac{x}{2}\geq& \left|\left|\frac{\sigma_0+\sigma_1}{2}-\sigma_{cl}\right|\right|_2\\
			\geq & \min_{\sigma_{cl}\in C}\left|\left|\frac{\sigma_0+\sigma_1}{2}-\sigma_{cl}\right|\right|_2
		\end{split}
		\label{gdsigma}
	\end{align}
	Using the definition of geometric discord (\ref{discord_general}) for the state $\frac{\sigma_0+\sigma_1}{2}$ and simplifying (\ref{gdsigma}), we get
\begin{align}
			x\geq& 2\sqrt{D_A^{(d,d)}\left(\frac{\sigma_0+\sigma_1}{2}\right)}\\
			x\log_2 x\geq& 2\sqrt{D_A^{(d,d)}\left(\frac{\sigma_0+\sigma_1}{2}\right)}\log_2 \left(2\sqrt{D_A^{(d,d)}\left(\frac{\sigma_0+\sigma_1}{2}\right)}\right)
		\end{align}
		Similarly, for the state $\frac{\sigma_{2}+\sigma_{3}}{2}$, we have
		\begin{align}
			z\geq& 2\sqrt{D_A^{(d,d)}\left(\frac{\sigma_2+\sigma_3}{2}\right)}\\
			z\log_2 z\geq& 2\sqrt{D_A^{(d,d)}\left(\frac{\sigma_2+\sigma_3}{2}\right)} \log_2 \left(2\sqrt{D_A^{(d,d)}\left(\frac{\sigma_2+\sigma_3}{2}\right)}\right)
	\end{align}
	Therefore, the distillable key rate $K_D$ can be calculated as 
	\begin{align}
		\begin{split}
			K_D=&1+x\log_2x+y\log_2y+z\log_2z+w\log_2w\\
			=&1+x\log_2x+z\log_2z\\
			\geq& 1+ 2\sqrt{D_A^{(d,d)}\left(\frac{\sigma_0+\sigma_1}{2}\right)}\log_2\left(2\sqrt{D_A^{(d,d)}\left(\frac{\sigma_0+\sigma_1}{2}\right)}\right)\\
			&+2\sqrt{D_A^{(d,d)}\left(\frac{\sigma_2+\sigma_3}{2}\right)}\log_2\left(2\sqrt{D_A^{(d,d)}\left(\frac{\sigma_2+\sigma_3}{2}\right)}\right)\\ 
			=& 1+2D_{1} \log_2 2D_1+2D_{2} \log_2 2D_2
		\end{split}
		\label{kd_formula_discord}
	\end{align}
	In the second line of the expression of $K_D$, given in (\ref{kd_formula_discord}), we have used $0\log_20=0$. Also, we introduce the notation $D_{1}=\sqrt{D_A^{(d,d)}\left(\frac{\sigma_0+\sigma_1}{2}\right)}$ and $D_2=\sqrt{D_A^{(d,d)}\left(\frac{\sigma_2+\sigma_3}{2}\right)}$, for the convenience.\\
	It can be clearly seen from (\ref{kd_formula_discord}) that the lower bound of $K_D$ depends upon the geometric discord of the $d^2$ dimensional states, denoted by $\sigma_i$'s where $i=0,1,2,3$. One can observe that the lower bound of the key rate $K_D$ derived in (\ref{kd_formula_discord}) can be negative sometime and thus does not assure the successful key generation always. Therefore, our task will be to find the intervals of geometric discord $D_{1}$ and $D_{2}$ where $K_D$ is non-negative, i.e., we need to discover the intervals of $D_{1}$ and $D_{2}$ for which successful generation of the secret key is guaranteed. To achieve this, we put the restriction on $D_{1}$ and $D_{2}$ for which the following inequality holds
	\begin{align}
		1+2 D_1 \log_2 2D_1 +2D_2\log_2 2D_2&\geq 0
		\label{kdg0_1}
	\end{align}
	Without any loss of generality, let us assume that $D_1\leq D_2$. Then, the inequality (\ref{kdg0_1}) reduces to
	\begin{align}
		\begin{split}
			\frac{-1}{4}&\leq D_1 \log_2 2D_1\\
			2^{-\frac{1}{4}}&\leq (2D_1)^{D_1}
		\end{split}
		\label{d1interval}
	\end{align}
	Therefore, there exist states $\sigma_{0}$ and $\sigma_{1}$ for which the inequality (\ref{d1interval}) holds when 
	\begin{align}
		\begin{split}
			&D_1\in [0,0.125)\cup (0.25,1]\\
			\implies& D_A^{(d,d)}\left(\frac{\sigma_0+\sigma_1}{2}\right)\in [0,0.015625)\cup (0.0625,1]
			\label{da_interval}
		\end{split}
	\end{align}
	The distillable key rate depends upon the behavior of the states $\sigma_i$'s. Therefore, the question arises that whether $\sigma_i$'s chosen by Alice and Bob can be separable, PPTES or NPTES, such that the secret key is generated successfully between the two communicating parties. In the next step, we will consider a few examples in which the two cases (i) $K_D> 0$ (ii) $K_D< 0$ holds. This implies the fact that in the first case the secret key is distributed between Alice and Bob and in the second case, the two distant parties fail to generate the secure secret key.\\  	
	
	\textbf{Example 1:} Let us start with the state $\varrho_1\in B(\mathbb{C}^2\otimes \mathbb{C}^2\otimes \mathbb{C}^2 \otimes \mathbb{C}^2)$ of the form (\ref{shared state kd})
	\begin{align}
		\begin{split}
			\varrho_1=&\ket{\phi^+}\bra{\phi^+}\otimes \sigma_{0} +\ket{\phi^-}\bra{\phi^-}\otimes \sigma_{1} +\ket{\psi^+}\bra{\psi^+}\otimes \sigma_{2} +\ket{\psi^-}\bra{\psi^-}\otimes \sigma_3
		\end{split}
		\label{rhokd2t2}
	\end{align}
	where $\ket{\phi^\pm}=\frac{1}{\sqrt{2}}(\ket{00}\pm\ket{11})$ and $\ket{\psi^\pm}=\frac{1}{\sqrt{2}}(\ket{01}\pm \ket{10})$. Further we define
	\begin{align}
		\begin{split}
			\sigma_0=\begin{pmatrix}
				\frac{q}{2}	 & 0 & 0 & 0 \\
				0	& \frac{1-q}{2} & \frac{\sqrt{1-q}}{2} & 0 \\
				0& \frac{\sqrt{1-q}}{2} & \frac{1}{2} & 0  \\
				0	& 0 & 0 & 0
			\end{pmatrix};~&	\sigma_1=\begin{pmatrix}
				\frac{1}{2} & 0&0&\frac{\sqrt{1-q}}{2}\\
				0&q &0&0\\
				0&0&0&0\\
				\frac{\sqrt{1-q}}{2}&0&0&\frac{1-q}{2}
			\end{pmatrix};\\
			\sigma_2= \begin{pmatrix}
				\frac{1}{2}&0&0&\frac{\sqrt{1-r}}{2}\\
				0&0&0&0\\
				0&0&0&0\\
				\frac{\sqrt{1-r}}{2}&0&0&\frac{1}{2}
			\end{pmatrix};~&		\sigma_3=\begin{pmatrix}
				0&0&0&0\\
				0&\frac{1}{2}&0&0\\
				0&0&\frac{1}{2}&0\\
				0&0&0&0
			\end{pmatrix}
		\end{split}
	\end{align}
	The operators $\sigma_{i},i=0,1,2,3$ satisfy the properties $O_{j},j=1,2,3$ when $q\in (0,0.4]$ and $r\in(0,0.4]$ and thus they represent the $2\otimes 2$ density matrices. The operators $\sigma_0, \sigma_1$ and $\sigma_2$ are NPT entangled states and $\sigma_3$ is a separable state. These states are taken such that $\left|\left|\sigma_0+\sigma_1\right|\right|_1=\left|\left|\sigma_0-\sigma_1\right|\right|_1$ and $\left|\left|\sigma_2+\sigma_3\right|\right|_1=\left|\left|\sigma_2-\sigma_3\right|\right|_1$ (taken upto 5 decimal places) holds.\\
	We choose a classical state described by the density operator $\sigma_{cl}$ of the following form
	\begin{align}
		\sigma_{cl}=p_1(\ket{\psi_1}\bra{\psi_1}\otimes \rho_1)+(1-p_1)(\ket{\psi_2}\bra{\psi_2}\otimes \rho_2)
		\label{sigma_cl_ex1}
	\end{align} 
	where $\ket{\psi_1}=\alpha\ket{0}+\beta\ket{1}$ and $\ket{\psi_2}=\alpha\ket{1}-\beta\ket{0}$, $|\alpha|^2+|\beta|^2=1,~\alpha,\beta\in \mathbb{C}$. Further, $\rho_k=\frac{1}{2}(I_2+\sum_{i=x,y,z} r_{ki}\lambda_i)$, $k=1,2$ with $\sum_i r_{ki}^{2}\leq 1$. $\lambda_i$ denotes the Pauli matrices for $i=x,y,z$, and $I_2$ denotes the identity matrix of order 2. \\
	For $r_{1x}=r_{1y}=r_{2x}=r_{2y}=\sqrt{0.1}$ and $r_{1z}=r_{2z}=0$; $\alpha=0.345$, $\beta=0.655$ and $p_{1}=0.24$, we find that there exist a classical state $\sigma_{cl}$ for which the inequalities $\left|\left|\frac{\sigma_0+\sigma_1}{2}\right|\right|_2\geq \left|\left|\frac{\sigma_0+\sigma_1}{2}-\sigma_{cl}\right|\right|_2$ and $\left|\left|\frac{\sigma_2+\sigma_3}{2}\right|\right|_2\geq \left|\left|\frac{\sigma_2+\sigma_3}{2}-\sigma_{cl}\right|\right|_2$ are satisfied. Therefore, we get
	\begin{align}
		\begin{split}	D_G^{(2,2)}\left(\frac{\sigma_0+\sigma_1}{2}\right)&=\frac{1-q}{16}\\
			D_G^{(2,2)}\left(\frac{\sigma_2+\sigma_3}{2}\right)&=\frac{1}{16}r^2
		\end{split}
	\end{align}
	Using the above values of $D_G^{(2,2)}\left(\frac{\sigma_0+\sigma_1}{2}\right)$ and $D_G^{(2,2)}\left(\frac{\sigma_2+\sigma_3}{2}\right)$, the lower bound of the distillable key rate $K_D$ for the state $\varrho_{1}$ can be calculated as 
	\begin{align}
		\begin{split}
			K_D&\geq 1+2\sqrt{D_G^{(2,2)}\left(\frac{\sigma_0+\sigma_1}{2}\right)}\log_2 2\sqrt{D_G^{(2,2)}\left(\frac{\sigma_0+\sigma_1}{2}\right)}\\
			&+2\sqrt{D_G^{(2,2)}\left(\frac{\sigma_2+\sigma_3}{2}\right)}\log_2 2\sqrt{D_G^{(2,2)}\left(\frac{\sigma_2+\sigma_3}{2}\right)}
		\end{split}
	\end{align}
	
	\begin{figure}[ht]
		\centering
		\includegraphics[width=0.7\linewidth]{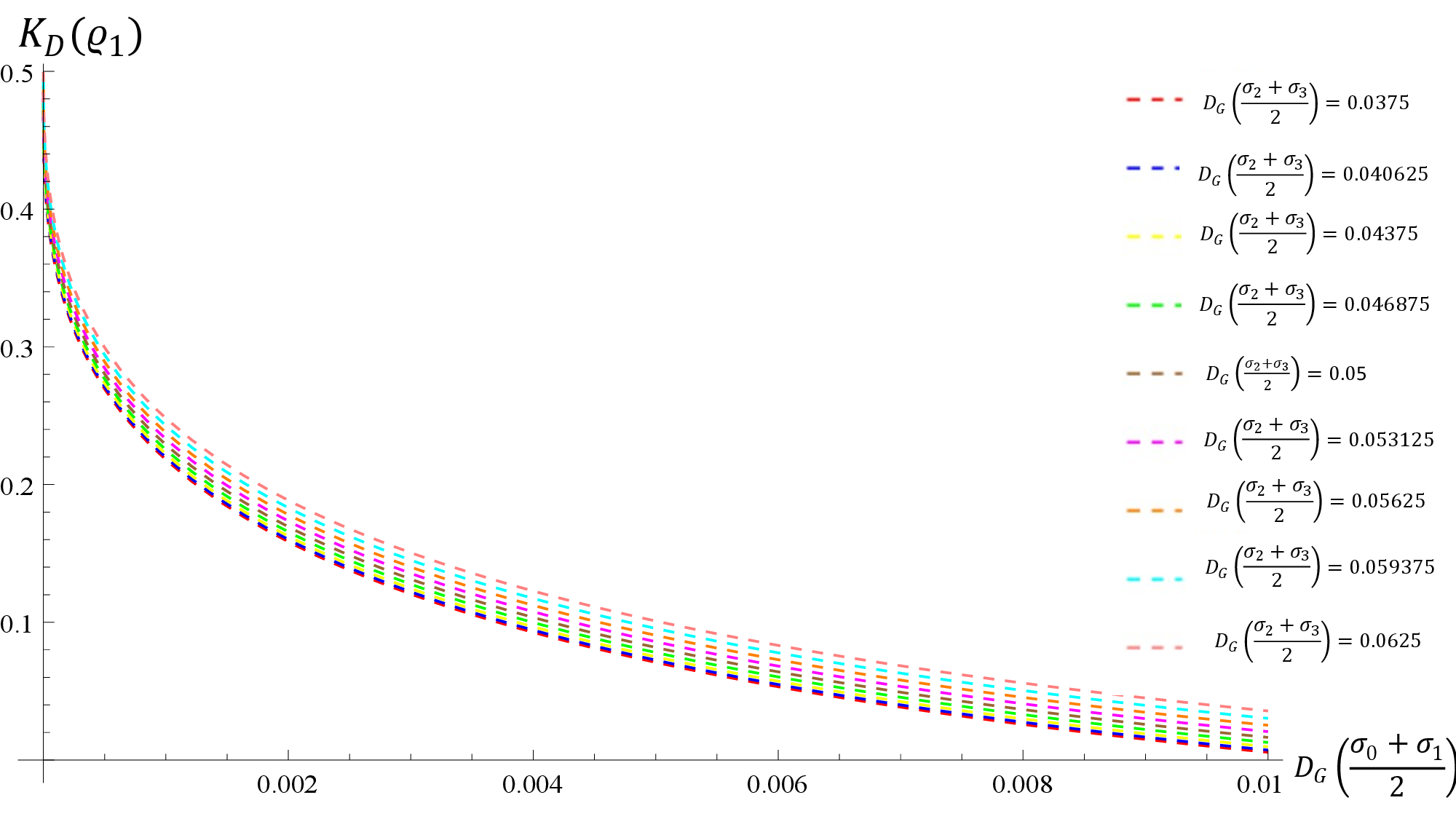}
		\caption{Distillable key rate $K_D$ for the state $\varrho_1$ defined in (\ref{rhokd2t2}) with $q\in (0,0.4]$ and $r\in (0,0.4]$. The Y-axis represents the distillable key rate $K_D(\varrho_1)$ and X-axis represents the range of $D_G^{(2,2)}\left(\frac{\sigma_0+\sigma_1}{2}\right)$. The graph is plotted corresponding to different values of $D_G^{(2,2)}\left(\frac{\sigma_2+\sigma_3}{2}\right)$.}
		\label{fig_kd2t2}
	\end{figure}
	The following insights can be drawn from the graph obtained in FIG.(\ref{fig_kd2t2}):\\
	(i) For some fixed value of $D_G^{(2,2)}\left(\frac{\sigma_2+\sigma_3}{2}\right)$, which belongs to the interval $(0.0375,0.0625]$, we find that, as the value of $D_G^{(2,2)}\left(\frac{\sigma_0+\sigma_1}{2}\right)\in (0,0.01]$ increases, the lower bound of $K_D(\varrho_1)$ decreases. \\
	(ii) For some fixed value of $D_G^{(2,2)}\left(\frac{\sigma_0+\sigma_1}{2}\right)$ which lies in the interval $(0,0.01]$, we observe that the  lower bound of distillable key rate increases as $D_G^{(2,2)}\left(\frac{\sigma_2+\sigma_3}{2}\right)\in (0.0375,0.0625]$ increases.\\
	Therefore, the value of $K_D$ is positive for this example.\\
	
\begin{figure}[ht]
		\centering
		\includegraphics[width=0.7\linewidth]{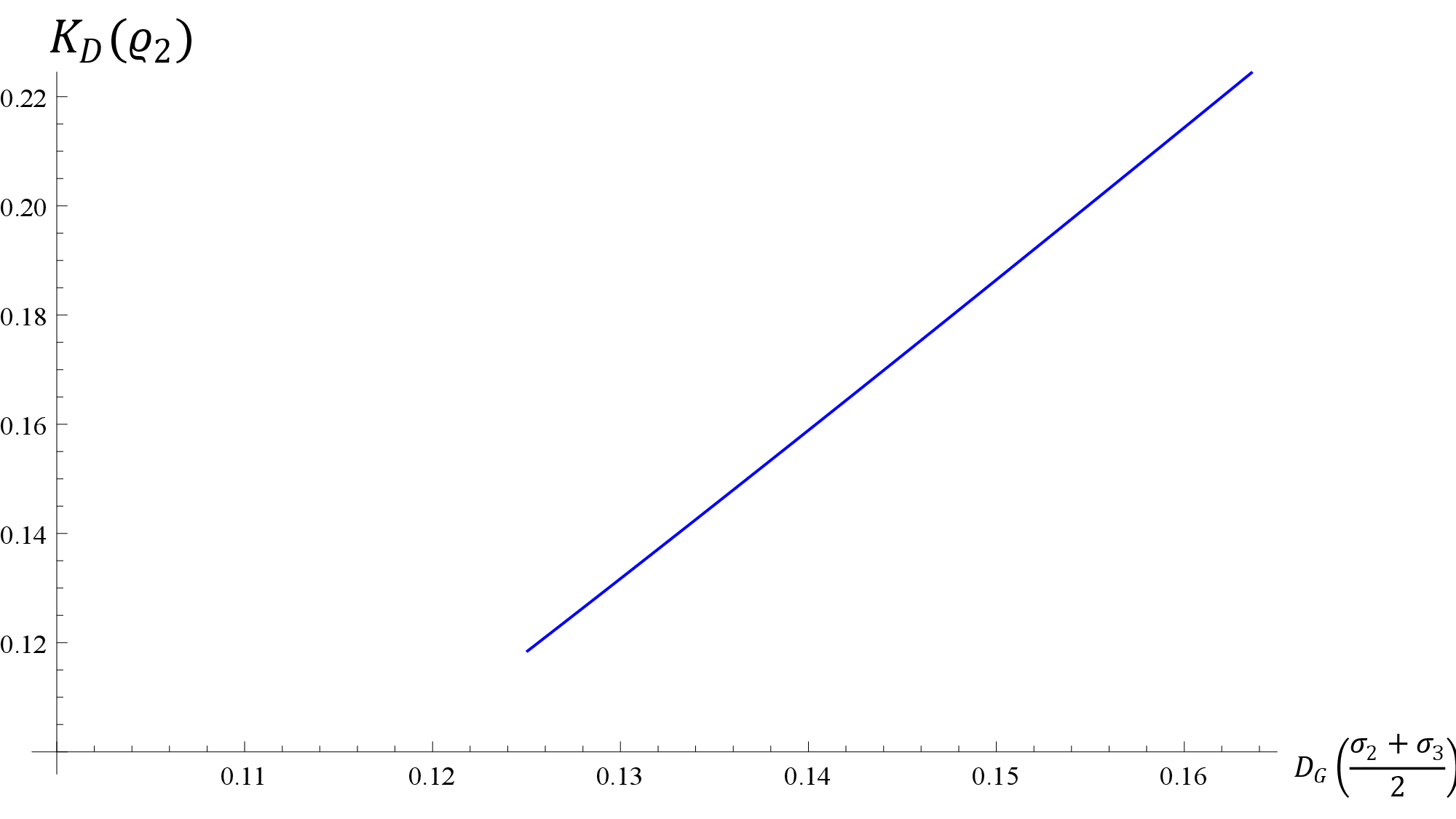}
		\caption{Distillable key rate $K_D$ for the state $\varrho_2$ defined in (\ref{rhokd2t2sep}) where $D_G^{(2,2)}\left(\frac{\sigma_2+\sigma_3}{2}\right)\in[0.125,0.163627]$. The X-axis represents $D_G^{(2,2)}\left(\frac{\sigma_2+\sigma_3}{2}\right)$ and Y-axis represents the corresponding distillable key rate $K_D(\varrho_2)$.}
		\label{fig_kd2t2sep}
	\end{figure}

	\textbf{Example 2:} Let us choose the state $\varrho_2\in B(\mathbb{C}^2\otimes \mathbb{C}^2\otimes \mathbb{C}^2 \otimes \mathbb{C}^2)$, which can be given by
	\begin{align}
		\begin{split}
			\varrho_2=&\ket{\phi^+}\bra{\phi^+}\otimes \sigma_{0} +\ket{\phi^-}\bra{\phi^-}\otimes \sigma_{1} +\ket{\psi^+}\bra{\psi^+}\otimes \sigma_{2} +\ket{\psi^-}\bra{\psi^-}\otimes \sigma_3
		\end{split}
		\label{rhokd2t2sep}
	\end{align}
	In this case, let us take
	\begin{align}
		\begin{split}
			\sigma_0=\begin{pmatrix}
				0.5 & 0.25 & 0.25 & 0 \\
				0.25	& 0.25 & 0& 0 \\
				0.25& 0 & 0.25 & 0  \\
				0	& 0 & 0 & 0
			\end{pmatrix};~&	\sigma_1=\begin{pmatrix}
				0 & 0&0&0\\
				0&0 &0&0\\
				0&0&0&0\\
				0&0&0&1
			\end{pmatrix};\\
			\sigma_2= \begin{pmatrix}
				\frac{m}{2}&0&\frac{\sqrt{1-m}}{2}&0\\
				0&1-m&0&0\\
				\frac{\sqrt{1-m}}{2}&0&1&0\\
				0&0&0&0
			\end{pmatrix};~&		\sigma_3=\begin{pmatrix}
				0 & 0&0&0\\
				0&0 &0&0\\
				0&0&0&0\\
				0&0&0&1
			\end{pmatrix}
		\end{split}
	\end{align}
	where $m\in[0,1]$. It can be observed here that the operators $\sigma_0$ and $\sigma_2$ are PPT entangled states, whereas the operators $\sigma_1$ and $\sigma_3$ are separable states. These operators satisfies the condition that $\left|\left|\sigma_0+\sigma_1\right|\right|_1=\left|\left|\sigma_0-\sigma_1\right|\right|_1$ and $\left|\left|\sigma_2+\sigma_3\right|\right|_1=\left|\left|\sigma_2-\sigma_3\right|\right|_1$.\\
	Let us consider the classical state $\sigma_{cl}$ as given in (\ref{sigma_cl_ex1}). We find that, for $r_{1x}=r_{1y}=r_{2x}=r_{2y}=\sqrt{0.1}$ and $r_{1z}=r_{2z}=0$; $\alpha=0.345$, $\beta=0.655$ and $p_{1}=0.24$, there exist a classical state $\sigma_{cl}$ for which the inequalities $\left|\left|\frac{\sigma_0+\sigma_1}{2}\right|\right|_2\geq \left|\left|\frac{\sigma_0+\sigma_1}{2}-\sigma_{cl}\right|\right|_2$ and $\left|\left|\frac{\sigma_2+\sigma_3}{2}\right|\right|_2\geq \left|\left|\frac{\sigma_2+\sigma_3}{2}-\sigma_{cl}\right|\right|_2$ holds. Thus, we have
	\begin{align}
		\begin{split}
			D_G^{(2,2)}\bigg(\frac{\sigma_0+\sigma_1}{2}\bigg)&=0.0275614\\
			D_G^{(2,2)}\bigg(\frac{\sigma_2+\sigma_3}{2}\bigg)&=\frac{1}{32}(3+(-2+m)m+\sqrt{m_1})
		\end{split}
	\end{align}
	where $m_1=5+m(-4+2m+m^3)$. Therefore, the distillable secret key rate $K_D$ can be illustrated as given in FIG.(\ref{fig_kd2t2sep}).
	
	From FIG.(\ref{fig_kd2t2sep}) it can be observed that for an increasing value of $D_G^{(2,2)}\left(\frac{\sigma_2+\sigma_3}{2}\right)$, the lower bound of the distillable key rate $K_D(\varrho_2)$ increases and is always positive for the given state $\varrho_2$. It is important to note here that even though the density operators $\sigma_i$'s are taken such that they are either PPTES or separable states, the distillable key can still be generated successfully. \\
	
	\textbf{Example 3:} Let us consider the state $\varrho_3\in B(\mathbb{C}^2\otimes \mathbb{C}^2\otimes \mathbb{C}^3 \otimes \mathbb{C}^3)$ defined as
	\begin{align}
		\begin{split}
			\varrho_3=&\ket{\phi^+}\bra{\phi^+}\otimes \sigma_{0} +\ket{\phi^-}\bra{\phi^-}\otimes \sigma_{1} +\\
			&\ket{\psi^+}\bra{\psi^+}\otimes \sigma_{2} +\ket{\psi^-}\bra{\psi^-}\otimes \sigma_3
		\end{split}
		\label{rhokd3t3}
	\end{align}
	where $\ket{\phi^\pm}$ and $\ket{\psi^\pm}$ are the two-qubit Bell states, and 
	\begin{align}
		\begin{split}
			\sigma_0=&\ket{\Psi}\bra{\Psi};\\
			\sigma_1=&\frac{1}{8}(I_9-\ket{\Psi}\bra{\Psi});\\
			\sigma_2=&\frac{1}{2}(\ket{01}\bra{01}+\ket{11}\bra{01}+\ket{01}\bra{11}+\ket{11}\bra{11});\\
			\sigma_3=&\frac{1}{5}\ket{\eta}\bra{\eta}
		\end{split}
	\end{align}
	are the $3\otimes 3$ density matrices, where $\ket{\Psi}=\frac{1}{\sqrt{3}}(\ket{00}+\ket{11}+\ket{22})$ and $\ket{\eta}=\ket{00}+\ket{02}+\ket{10}+\ket{12}+\ket{22}$. The states $\sigma_0$, $\sigma_1$, $\sigma_2$ and $\sigma_3$ are chosen in such a way that $O_{4}$ is satisfied.\\
	The geometric discord of the state $\varrho_3$ is given by $  D_G(\varrho_3)=0.000931385$. Now, let us consider the zero discord state $\sigma_{cl}$ is of the form
	\begin{align}
		\sigma_{cl}=p_1 \ket{\psi_1}\bra{\psi_1} \otimes \rho_1 +p_2 \ket{\psi_2}\bra{\psi_2} \otimes \rho_2 +p_3\ket{\psi_3}\bra{\psi_3} \otimes \rho_3
	\end{align}
	where $\ket{\psi_k}, ~k=1,2,3$ denote the orthogonal vectors given in the form
		\begin{align}
			\begin{split}
				\ket{\psi_1}&=\alpha \ket{0}+\beta \ket{1}+\gamma \ket{2}\\
				\ket{\psi_2}&=\frac{-\beta \ket{0}+\alpha \ket{1}}{\sqrt{\alpha^2+\beta^2}}\\
				\ket{\psi_3}&=\frac{-\alpha \gamma \ket{0}-\beta \gamma \ket{1}+(\alpha^2+\beta^2)\ket{2}}{\sqrt{\alpha^2+\beta^2}}
			\end{split}
		\end{align}
		with $p_1+p_2+p_3=1$. Further, $\rho_k=\frac{1}{3}I_3+\sum_{i}b_{ki}\Lambda^{(i)}$ for $k=1,2,3$ and $i=1,2,\dots, 8$. $\Lambda^{(i)}$ denotes the Gell-Mann matrices of order 3 defined in (\ref{gell_mann matrices}), and $I_3$ denote the identity matrix of order 3.\\
	There exists a classical state $\sigma_{cl}$ for $b_{11}=b_{21}=b_{31}=b_{12}=b_{22}=b_{32}=b_{18}=\sqrt{0.1}$, $b_{14}=b_{24}=b_{34}=\sqrt{0.2}$, $b_{16}=b_{36}=\sqrt{0.5}$, $b_{28}=\sqrt{0.3}$, $p_1=0.084$, $p_2=0.866$, $p_3=0.05$, $\alpha=0.031$, $\beta=0.656$ and $\gamma=\sqrt{1-\alpha^2-\beta^2}$, such that $\left|\left|\frac{\sigma_0+\sigma_1}{2}\right|\right|_2\geq \left|\left|\frac{\sigma_0+\sigma_1}{2}-\sigma_{cl}\right|\right|_2$ and $\left|\left|\frac{\sigma_2+\sigma_3}{2}\right|\right|_2\geq \left|\left|\frac{\sigma_2+\sigma_3}{2}-\sigma_{cl}\right|\right|_2$ holds. 
	Therefore, we have
	\begin{align}
		\begin{split}
			D_G^{(3,3)}\left(\frac{\sigma_0+\sigma_1}{2}\right)=&0.0252058\\
			D_G^{(3,3)}\left(\frac{\sigma_2+\sigma_3}{2}\right)=&0.0848299\\
		\end{split}
		\label{da3_example2}
	\end{align}
	Using (\ref{da3_example2}), we can calculate the value of the lower bound of the distillable key rate $K_D$ for the state $\varrho_3$, given by
	\begin{align}
		\begin{split}
			K_D\geq& 1+2\sqrt{0.0252058}\log_2 2\sqrt{0.0252058}\\
			&+2\sqrt{0.0848299}\log_2 2\sqrt{0.0848299}\\
			=&0.0203274
		\end{split}
	\end{align}
	We observe that $K_D>0$, which implies that distillation of the secret key is possible in the case when Alice and Bob share the five qubit state $\varrho_3$ given by (\ref{rhokd3t3}).\\
	
	\textbf{Example 4:} In this example, we consider a state $\varrho_4\in B(\mathbb{C}^2\otimes \mathbb{C}^2 \otimes \mathbb{C}^2 \otimes \mathbb{C}^2)$ with geometric discord $D_G(\varrho_4)=0.000747035$. Here, $\sigma_{i},i=0,1,2,3$ can be expressed as
	\begin{align}
		\begin{split}
			\sigma_0=\begin{pmatrix}
				0.5 & 0 & 0 & 0.25 \\
				0 & 0 & 0 & 0 \\
				0 & 0 & 0 & 0 \\
				0.25 & 0 & 0 & 0.5
			\end{pmatrix};~ & \sigma_1=\begin{pmatrix}
				0 & 0 & 0 & 0 \\
				0 & 0.5 & 0.5 & 0\\
				0 & 0.5 & 0.5 & 0 \\
				0 & 0 & 0 & 0
			\end{pmatrix};\\
			\sigma_2=\begin{pmatrix}
				0.5 & 0 & 0.5 & 0 \\
				0 & 0 & 0 & 0 \\
				0.5 & 0 & 0.5 & 0 \\
				0 & 0 & 0 & 0
			\end{pmatrix};~& \sigma_3=\begin{pmatrix}
				0 & 0 & 0 & 0 \\
				0 & 0.25 & 0 & 0.1 \\
				0 & 0 & 0 & 0 \\
				0 & 0.1 & 0 & 0.75
			\end{pmatrix}
		\end{split}  
	\end{align}
	We find that $D_G^{(2,2)}(\frac{\sigma_0+\sigma_1}{2})=0.015625$ and $D_G^{(2,2)}(\frac{\sigma_2+\sigma_3}{2})=0.041971$. Upon substituting these values in (\ref{kd_formula_discord}), we find that
	\begin{align}
		\begin{split}
			K_D\geq&1+(2\sqrt{0.015625}) \log_2 (2\sqrt{0.015625})\\
			&+(2\sqrt{0.041971}) \log_2 (2\sqrt{0.041971})\\
			=&-0.0274266
		\end{split}
	\end{align}
	Therefore, it can be noted that if $D_G^{(2,2)}(\frac{\sigma_0+\sigma_1}{2})$ take the value $0.015625$ then the value of $K_{D}$ may be negative. Thus, the range of $D_G^{(2,2)}(\frac{\sigma_0+\sigma_1}{2})$ given in (\ref{da_interval})  for $K_{D}>0$ is verified. 
	
	\section{Comparison with the existing works}
	There are many works, for the development of discord based QKD schemes, available in the literature \cite{dakic_2010, wang_2024,luo_2008,wu_2015,kim_2009,sterltsov_2013,debarba_2012}. We, therefore, compare	our QKD protocol with the others, which are as given below
		\begin{table}[ht]
			\begin{tabular}{|p{0.5\linewidth}|p{0.5\linewidth}|}
				\hline
				\multicolumn{1}{|c|}{Existing Works} & \multicolumn{1}{c|}{Our Work} \\ \hline
				Dakic et al \cite{dakic_2010} introduced the idea of geometric discord with a closed form expression for $2\otimes 2$ systems. They did not provide any links with QKD and secure key rates. & In our work, we introduce the concept of geometric discord in arbitrary dimensional bipartite systems along with its applications in QKD. \\ \hline			
				Recently, in 2024, Wang et. al \cite{wang_2024} proved that even in the absence of entanglement, discord guarantees some level of secrecy against eavesdropping in qubit based protocols. They provided a conceptual link between quantum discord and the secret key using quantum discord. & 
				Calculation of quantum discord in higher dimensional systems might be complex \cite{luo_2008}. Therefore, in our work, we adopt the idea of using geometric quantum discord. We show how GQD contributes to private communication and key extraction in the presence of Eve. We have established an explicit relation between GQD and the distillable key rate $K_D$. \\ \hline
				Wu and Zhou, in 2013 studied geometric discord as a resource for improving QKD rates via twirling operations, focusing on two-qubit systems \cite{wu_2015}. 
				& We provide a general idea on how geometric discord contributes to distillable key rates even in the presence of Eve, without relying on specific twirling operations. \\ \hline
				Kim and Sanders \cite{kim_2009} and Streltsov et al \cite{sterltsov_2013} showed that entanglement is not an compulsory requirement for secret key generation. Their works did not consider other correlations measured by discord and its contribution to the generation of secret key. & Our work highlights the usefulness of geometric discord in successful generation of the secret key between two communicating parties, which goes beyond the limited scope of entanglement in previous works. \\
				\hline
				Most works focuses upon providing entanglement based bounds or numerical simulations for the same \cite{debarba_2012}.	& We derive a lower bound for the distillable secret key rate $K_D$ in terms of geometric quantum discord providing an analytical tool for evaluation of the security of QKD protocols for arbitrary dimensional bipartite systems.\\
				\hline
			\end{tabular}
		\end{table}	
	
	\section{Conclusion}
	To summarize, we have derived the analytical expression of geometric discord for a two-qutrit ($3\otimes 3$) system, and then generalized the analytical expression of geometric discord applicable to $d_1\otimes d_2$ dimensional system. Using the derived analytical formula for GQD, we analyzed a few examples of $3\otimes 3$ and $4\otimes 4$ dimensional systems and observed that GQD captures the presence of quantum correlations. For the states with no correlations, the GQD is zero. In case of separable states, the non-zero discord indicates the presence of quantum correlations.\\
	Since geometric discord quantifies quantum correlations, we tried to see whether there exist any relationship between GQD and the amount of entanglement contained in a state measured by negativity. We found that there is a definite relationship between negativity and GQD for a given state but that relationship may not hold for all quantum states. With the help of a few examples, we found that GQD may be greater than or less than the negativity for a quantum state under consideration. Further, we use the concept of geometric discord in studying the QKD protocols. In particular, we focus upon finding the lower bound for the distillable key rate $K_D$ in terms of GQD $D_G^{(d,d)}(\rho)$ when two communicating parties make use of private state $\sigma_{AB}$ for generating a secret key in the presence of an eavesdropper. We find that when $\sqrt{D_A^{(d,d)}\left(\frac{\sigma_0+\sigma_1}{2}\right)}\in [0.125,0.25]$, then $K_D<0$ and therefore the successful generation of the secret key is not guaranteed, where $\sigma_0$ and $\sigma_1$ represents the $d^2$ dimensional density matrices. The choice of $\sigma_0$ and $\sigma_1$ depends upon the communicating parties, and they may select them in such a way that $K_D>0$. We then discuss a few examples for both the cases when (i) $K_D>0$, i.e., the secret key is distributed between the communicating parties and (ii) $K_D<0$, i.e., the two parties fail to successfully generate a secret key. We observe that even if $\sigma_i$'s are taken as PPTES or separable states, the distillable key can still be generated. \\
	The idea of using geometric discord in a QKD protocol can offer a more effective approach because of the following reasons. First, 
		our results reveal that generation of a secure key is possible even in the absence of entanglement, suggesting that the correlations beyond entanglement plays a crucial role in successful generation of the secret key. This result highlights GQD as a valuable resource for QKD and expands the possibilities for secure quantum communications. Second, the derived lower bound of $K_D$ ensures the successful generation of secret key between two communicating parties using private states in certain intervals of $D_A^{(d,d)}\left(\frac{\sigma_0+\sigma_1}{2}\right)$, which we found were missing in earlier work \cite{chi_2007}. \\
In future work, we would like to study the performance of GQD based QKD protocols under more realistic noisy quantum channels such as amplitude damping, phase damping and depolarizing channels. In addition to this, it would also be interesting to extend the presented work to multi-party scenario.

	\section{Author contribution statement}
	\textbf{RJ, SA}: Conceptualization, Methodology, Software, Writing- Original draft preparation, Visualization, Writing- Reviewing and Editing.
	
	\section{Competing interests}
	The authors have no competing interests to declare that are relevant to the content of this article.
		
                    \balance

					\section{Appendix}
					\begin{appendices}

	\section{Determination of triplet $\{\vb*{a_1^{(i)}}t_1+\vb*{a_2^{(i)}}t_2, \vb*{u}, S\}$ of the $3 \otimes 3$ dimensional state $\chi_{AB}^{3\otimes 3}$}
	The $3 \otimes 3$ dimensional zero-discord state $\chi_{AB}^{3\otimes 3}$ can be written as
	\begin{align}
		\chi_{AB}^{3\otimes 3}=p_1 \ket{\psi_1}\bra{\psi_1} \otimes \rho_1 +p_2 \ket{\psi_2}\bra{\psi_2} \otimes \rho_2 + p_3 \ket{\psi_3}\bra{\psi_3} \otimes \rho_3
	\end{align}
	where $p_i\geq0,~(i=1,2,3)$ such that $\sum_{i=1}^3p_i=1$ and $\rho_k, ~ k=1,2,3$ denotes the single qutrit generalized density matrices, given in (\ref{rho_k}).
	$\{\ket{\psi_1},\ket{\psi_2},\ket{\psi_3}\}$ are single-qubit states given in (\ref{psi}).
	Similar to the triplet of the state $\rho=\{\vb*{x},\vb*{y},T\}$, the state $\chi_{AB}^{3\otimes 3}$ can also be expressed in a triplet $\{\vb*{a_1^{(i)}}t_1+\vb*{a_2^{(i)}}t_2, \vb*{u}, S\}$. Therefore, our task is now to determine the components of $\vb*{a_1^{(i)}}t_1+\vb*{a_2^{(i)}}t_2$, the component of  $\vb*{u}$ and the correlation matrix $S$.
	
	\subsection{Calculation of the components of $\vb*{a_1^{(i)}}t_1+\vb*{a_2^{(i)}}t_2$}
	Let us start with $\text{Tr}\left[\chi(\Lambda^{(i)}\otimes I_3)\right]$, which represent the $i^{th}$ component of the local Bloch vector of system $A$. It is given by 
	\begin{align}
		\begin{split}
			\text{Tr}\left[\chi(\Lambda^{(i)}\otimes I_3)\right]=&p_1 \bra{\psi_1}\Lambda^{(i)}\ket{\psi_1}+p_2 \bra{\psi_2}\Lambda^{(i)}\ket{\psi_2} \\
			&+p_3 \bra{\psi_3}\Lambda^{(i)}\ket{\psi_3},~i=1,2,\dots,8
			\label{trchix}
		\end{split}
	\end{align}
	Using the completeness relation $\ket{\psi_1}\bra{\psi_1}+\ket{\psi_2}\bra{\psi_2}+\ket{\psi_3}\bra{\psi_3}=I_3$, we have
	\begin{align}
		\begin{split}
			\bra{\psi_3}\Lambda^{(i)}\ket{\psi_3}=-\bra{\psi_1}\Lambda^{(i)}\ket{\psi_1}-\bra{\psi_2}&\Lambda^{(i)}\ket{\psi_2},~i=1,2,\dots,8
		\end{split}
		\label{trchix_prop}
	\end{align}
	Let $a_k^{(i)}=\bra{\psi_k}\Lambda^{(i)}\ket{\psi_k}$ for $i=1,2,\dots, 8$ and $k=1,2$. \\
	Using (\ref{trchix_prop}) in (\ref{trchix}), we get
	\begin{align}
		\begin{split}
			\text{Tr}&\left[\chi(\Lambda^{(i)}\otimes I_3)\right]=p_1 a_1^{(i)}+ +p_2a_2^{(i)}+p_3a_3(a_1^{(i)}-a_2^{(i)})\\
			=&a_1^{(i)} (p_1-p_3)+a_2^{(i)}(p_2-p_3),~i=1,2,\dots,8
		\end{split}
	\end{align}
	Now, let us substitute $t_1=p_1-p_3$ and $t_2=p_2-p_3$ and we get for each $i=1,2,\dots,8$
	\begin{align}
		\text{Tr}\left[\chi(\Lambda^{(i)}\otimes I_3)\right]=& a_1^{(i)}t_1+a_2^{(i)}t_2\equiv \sum_{k=1}^2 a_k^{(i)} t_k
		\label{trchix_fin}
	\end{align}
	Now it can be seen that $a_1^{(i)}$ and $a_2^{(i)}$ take eight values when $i=1,2,\dots,8$. Therefore, $a_1^{(i)}$ and $a_2^{(i)}$ can be considered as a eight-component vector which can be expressed as $\vb*{a_k^{(i)}}=(a_k^{(1)},a_k^{(2)},\dots, a_k^{(8)})^T,~k=1,2$.	Using (\ref{psi}), the component of the vectors $\vb*{a_1^{(i)}}$ and $\vb*{a_2^{(i)}}$ can be calculated as
	\begin{align}
		\begin{split}
			\vb*{a_1^{(i)}}&= \bigg(
			2 \alpha \beta~ 2 \alpha \gamma ~2 \beta \gamma	~0~0~0~\alpha^2-\beta^2~ \frac{1}{\sqrt{3}}(\alpha^2+\beta^2-2\gamma^2)\bigg)^T\\		
			\vb*{a_2^{(i)}}&=\bigg(\frac{-2 \alpha \beta}{\alpha^2+\beta^2}~0~0~0~0	~0~	\frac{\beta^2-\alpha^2}{\alpha^2+\beta^2}	~\frac{1}{\sqrt{3}}\bigg)^T\\
		\end{split}
		\label{a_k^{(i)} vectors}
	\end{align}
	Thus, using (\ref{a_k^{(i)} vectors}), $\vb*{a_1^{(i)}}t_1+\vb*{a_2^{(i)}}t_2$ can be calculated as
	
	\begin{align}
		\begin{split}
			\vb*{a_1^{(i)}}t_1+\vb*{a_2^{(i)}}t_2	=&\begin{pmatrix}
				2\alpha \beta \left(t_1-\frac{t_2}{\alpha^2+\beta^2}\right)	\\
				2 \alpha \gamma t_1	\\
				2 \beta \gamma t_1	\\
				0	\\
				0	\\
				0	\\
				(\alpha^2-\beta^2) \left(t_1-\frac{t_2}{\alpha^2+\beta^2}\right)	\\
				\frac{1}{\sqrt{3}}\left((\alpha^2+\beta^2-2\gamma^2)t_1+t_2\right)
			\end{pmatrix}
		\end{split}
	\end{align}

	\subsection{Calculation of the components of $\vb*{u}$}
	Let us now begin by considering $\text{Tr}\left[\chi(I_3 \otimes \Lambda^{(i)})\right]$ which represent the $i^{th}$ component of the local Bloch vector of system $B$. It is given by 
	\begin{align}
		\begin{split}
			\text{Tr}\left[\chi(I_3 \otimes \Lambda^{(i)})\right]
			&=\text{Tr}\left[(p_1\rho_1+p_2\rho_2+p_3\rho_3)\Lambda^{(i)}\right]
			\label{trchiy}
		\end{split}
	\end{align}
	Let $u_i=\text{Tr}\left[(p_1\rho_1+p_2\rho_2+p_3\rho_3)\Lambda^{(i)}\right]$ represent $i^{th}$ component of a eight-component vector (say $\vb*{u}$). Therefore, we find that
	\begin{align}
		\begin{split}
			u_i=& \text{Tr}\left[(p_1\rho_1+p_2\rho_2+p_3\rho_3)\Lambda^{(i)}\right]\\
			=&p_1 \text{Tr}(\rho_1 \Lambda^{(i)})+p_2 \text{Tr}(\rho_2 \Lambda^{(i)})+p_3 \text{Tr}(\rho_3 \Lambda^{(i)})\\
			=& p_1 b_i +p_2 c_i +p_3 d_i
		\end{split}
	\end{align}
	where $b_i=\frac{1}{2}\text{Tr}[\rho_1 \Lambda^{(i)}]$, $c_i=\frac{1}{2}\text{Tr}[\rho_2 \Lambda^{(i)}]$ and $d_i=\frac{1}{2}\text{Tr}[\rho_3 \Lambda^{(i)}], ~i=1,2,\dots, 8$.
	Therefore, we obtain
	\begin{align}
		\vb*{u}= \begin{pmatrix}
			p_1 b_1+p_2 c_1+p_3 d_1	\\
			p_1 b_2+p_2 c_2+p_3 d_2		\\
			p_1 b_3+p_2 c_3+p_3 d_3		\\
			p_1 b_4+p_2 c_4+p_3 d_4		\\
			p_1 b_5+p_2 c_5+p_3 d_5		\\
			p_1 b_6+p_2 c_6+p_3 d_6		\\
			p_1 b_7+p_2 c_7+p_3 d_7		\\
			p_1 b_8+p_2 c_8+p_3 d_8
		\end{pmatrix}
	\end{align}
	
	\subsection{Calculation of the matrix element of the correlation matrix of $S$}
	Let us consider the $\text{Tr}\left[\chi(\Lambda^{(i)} \otimes \Lambda^{(j)})\right]$ which can be denoted as the $(i,j)^{th}$ term of the correlation matrix $S$. For all $i,j=1,2,\dots,8$, $\text{Tr}\left[\chi(\Lambda^{(i)} \otimes \Lambda^{(j)})\right]$ is given by
	\begin{align}
		\begin{split}
			S^{(i,j)}=\text{Tr}\left[\chi(\Lambda^{(i)} \otimes \Lambda^{(j)})\right]
			&=a_1^{(i)} \text{Tr}\left[(p_1\rho_1-p_3\rho_3)\Lambda^{(j)}\right]\\
			&+a_2^{(i)}\text{Tr}\left[(p_2\rho_2-p_3\rho_3)\Lambda^{(j)}\right]\\
			=&\sum_{k=1}^{2}a_k^{(i)} \text{Tr}[(p_k\rho_k-p_3\rho_k)\Lambda^{(j)}]
		\end{split}
		\label{trchit}
	\end{align}
	Using (\ref{rho_k}), (\ref{trchit}) reduces to
	\begin{small}
		\begin{align}
			\begin{split}
				S^{(i,j)}&=
				a_1^{(i)} (p_1 b_j-p_3 d_j)+a_2^{(i)} (p_2 c_j-p_3 d_j)
			\end{split}
		\end{align}
	\end{small}

	\section{Calculations for $\normsq{\rho_{AB}^{3\otimes 3}}_2$ and $\normsq{\chi_{AB}^{3\otimes 3}}_2$} 
	Let us start with the two-qutrit density matrix $\rho_{AB}^{3\otimes 3}$, defined as
	\begin{align}
		\begin{split}	
			\rho_{AB}^{3\otimes 3}=\frac{1}{9}&\bigg(I_3 \otimes I_3 +\sum_{i=1}^{8} x_i \Lambda^{(i)}\otimes I_3 +\sum_{i=1}^{8} y_i I_3 \otimes \Lambda^{(i)} +\sum_{i,j=1}^{8}T_{ij} \Lambda^{(i)}\otimes \Lambda ^{(j)}\bigg)
		\end{split}
		\label{rho_appendixb}
	\end{align}
	Since $\rho_{AB}^{3\otimes 3}$ is hermitian, therefore we have, $\rho_{AB}^{3\otimes 3}=\rho_{AB}^{(3\otimes 3)\dagger}$, thus, $\text{Tr}(\rho_{AB}^{(3\otimes 3)\dagger}\rho_{AB}^{3\otimes 3})= \text{Tr}(\rho_{AB}^{(3\otimes 3)2})$. 
	Using the properties of Gell-Mann matrices, (\ref{rho_appendixb}) reduces to
	\begin{align}
		\begin{split}
			\normsq{\rho_{AB}^{3\otimes 3}}_2=&=\frac{1}{81}\left(9+ 6 \sum_{i=1}^{8} x_i^2+6 \sum_{i=1}^{8} y_i^2+4 \sum_{i,j=1}^{8} T_{ij}^2\right)\\
			&=\frac{1}{81}\left(9+ 6 \normsq{\vb*{x}}_2+6\normsq{\vb*{y}}_2+4 \normsq{T_{ij}}\right)
		\end{split}
		\label{norm rho}
	\end{align}
	Proceeding analogously like $\normsq{\rho_{AB}^{3\otimes 3}}_2$, we can use the triplet form of the zero discord state $\chi_{AB}^{3\otimes 3}$ and then we can derive the value of $\normsq{\chi_{AB}^{3\otimes 3}}_2$ as follows.
	\begin{align}
		\normsq{\chi_{AB}^{3\otimes 3}}_2=\frac{1}{81}\left(9+ 6 \normsq{\vb*{a_1^{(i)}}t_1+\vb*{a_2^{(i)}}t_2}_2+6\normsq{\vb*{u}}_2+4\normsq{S}_2\right)
		\label{norm chi}
	\end{align}

	\section{Minimize $\normsq{\rho_{AB}^{3\otimes 3}-\chi_{AB}^{3\otimes 3}}_2$ over all zero discord states $\chi_{AB}$}
	
	To minimize $\normsq{\rho_{AB}^{3\otimes 3}-\chi_{AB}^{3\otimes 3}}_2$, we recall the expression of it from (\ref{norm_rho_chi}), and thus we have
	\begin{align}
		\begin{split}
			\normsq{\rho_{AB}^{3\otimes 3}-\chi_{AB}^{3\otimes 3}}_2&=\frac{2}{81} \left(3 \normsq{\vb*{x}}_2+3\normsq{\vb*{y}}_2+2 \normsq{T}\right)\\
			&+\frac{6}{81}\normsq{\vb*{u}}_2-\frac{12}{81}\vb*{y}\cdot\vb{u}+h+g
		\end{split}
		\label{rhomin_c1}
	\end{align}
	where 
	\begin{align}
		\begin{split}
			h=&\frac{1}{81}\bigg(8(t_1^2+t_2^2-t_1t_2)-12(t_1{\vb*{x}}^T\vb{a_1^{(i)}}+t_2{\vb*{x}}^T \vb*{a_2^{(i)}})\bigg)\\
			g=&\frac{4}{81} \bigg(\normsq{\vb*{a_1^{(i)}}{\vb*{r_1^{(j)}}}^T}_2-\frac{2}{3} {\vb*{r_1^{(j)}}}^T\vb{r_2^{(j)}}-\frac{2}{3} {\vb*{r_2^{(j)}}}^T\vb{r_1^{(j)}}+\\
			&\normsq{\vb*{a_2^{(i)}}{\vb*{r_2^{(j)}}}^T}_2\bigg)-\frac{8}{81} \bigg({\vb*{r_1^{(j)}}}^T T \vb*{a_1^{(i)}}+{\vb*{r_2^{(j)}}}^T T \vb*{a_2^{(i)}}\bigg)
		\end{split}
		\label{h_g_func}
	\end{align}
	where $\vb*{x}=(x_1~x_2~x_3~x_4~x_5~x_6~x_7~x_8)^T$,  $\vb*{y}=(y_1~y_2~y_3~y_4~y_5~y_6~y_7~y_8)^T$, and $\vb*{r_k^{(j)}}=(r_k^{(1)}~r_k^{(2)}~r_k^{(3)}~r_k^{(4)}~r_k^{(5)}~r_k^{(6)}~r_k^{(7)}~r_k^{(8)})^T$ and the components $r_k^{(j)}=\text{Tr}[(p_k\rho_k-p_3\rho_k)\Lambda^{(j)}]$ for $k=1,2, ~j=1,2,\dots,8$, $T$ is the correlation matrix of $\rho_{AB}^[3\otimes 3]$ of order $8\times 8$ and $\normsq{T}_2=\text{Tr}T^T T$. \\
	To minimize $\normsq{\rho_{AB}^{3\otimes 3}-\chi_{AB}^{3\otimes 3}}_2$, we calculate the partial derivative of $\normsq{\rho_{AB}^{3\otimes 3}-\chi_{AB}^{3\otimes 3}}_2$ with respect to the parameters $t_1,t_2,\vb*{u},\vb*{r_1^{(j)}}$ and $\vb*{r_2^{(j)}}$ and equate each partial derivative to zero.
	
	\begin{align}
		\begin{split}
			\frac{\partial\normsq{\rho_{AB}^{3\otimes 3}-\chi_{AB}^{3\otimes 3}}_2}{\partial t_1}=0
			\implies4t_1-2t_2=3{\vb*{x}}^T\vb{a_1^{(i)}}
		\end{split}
		\label{t1t2}
	\end{align}
	\begin{align}
		\begin{split}
			\frac{\partial\normsq{\rho_{AB}^{3\otimes 3}-\chi_{AB}^{3\otimes 3}}_2}{\partial t_2}=0
			\implies4t_2-2t_1=3{\vb*{x}}^T\vb{a_2^{(i)}}
		\end{split}
		\label{t2t1}
	\end{align}
	\begin{align}
		\begin{split}
			\frac{\partial \normsq{\rho_{AB}^{3\otimes 3}-\chi_{AB}^{3\otimes 3}}_2}{\partial \vb*{u}}=0
			\implies\vb{u}=\vb*{y}
		\end{split}
		\label{s+y}
	\end{align}
	\begin{align}
		\begin{split}
			\frac{\partial \normsq{\rho_{AB}^{3\otimes 3}-\chi_{AB}^{3\otimes 3}}_2}{\partial \vb*{r_1^{(j)}}}=0
			\implies \vb*{r_2^{(j)}}=\frac{3}{2}(\vb*{r_1^{(j)}}{\vb*{a_1^{(i)}}}^T\vb{a_1^{(i)}}-T\vb{a_1^{(i)}})
		\end{split}
		\label{s-2}
	\end{align}
	
	\begin{align}
		\begin{split}
			\frac{\partial \normsq{\rho_{AB}^{3\otimes 3}-\chi_{AB}^{3\otimes 3}}_2}{\partial \vb*{r_2^{(j)}}}=0
			\implies \vb*{r_1^{(j)}}=\frac{3}{2}(\vb*{r_2^{(j)}}{\vb*{a_2^{(i)}}}^T\vb{a_2^{(i)}}-T\vb{a_2^{(i)}})
		\end{split}
		\label{s-1}
	\end{align}
	Solving the equations (\ref{t1t2}) and (\ref{t2t1}), we get
	\begin{align}
		\begin{split}
			t_1=&{\vb*{x}}^T\vb{a_1^{(i)}}+\frac{1}{2}{\vb*{x}}^T\vb{a_2^{(i)}}\\
			t_2=&\frac{1}{2}{\vb*{x}}^T\vb{a_1^{(i)}}+{\vb*{x}}^T\vb{a_2^{(i)}}
		\end{split}
		\label{t1,t2 final}
	\end{align}
	Further, solving (\ref{s-2}) and (\ref{s-1}), we get
	\begin{align}
		\begin{split}
			\vb*{r_1^{(j)}}=&\frac{3}{4} T\left(\vb*{a_1^{(i)}}\left({\vb*{a_2^{(i)}}}^T\vb{a_2^{(i)}}\right)+\frac{2}{3}\vb*{a_2^{(i)}}\right)\\
			\vb*{r_2^{(j)}}=&\frac{3}{4} T\left(\vb*{a_2^{(i)}}\left({\vb*{a_1^{(i)}}}^T\vb{a_1^{(i)}}\right)+\frac{2}{3}\vb*{a_1^{(i)}}\right)
		\end{split}
		\label{s-12 final}
	\end{align}
	Using (\ref{s+y}), (\ref{t1,t2 final}) and (\ref{s-12 final}) in (\ref{rhomin_c1}) and (\ref{h_g_func}), we get
	\begin{align}
		\begin{split}
			\normsq{\rho_{AB}^{3\otimes 3}-\chi_{AB}^{3\otimes 3}}_2&=\frac{2}{81}\big(3\normsq{\vb*{x}}_2+2\normsq{T}_2\big)\\
			&-\frac{2}{81}\bigg[{\vb*{a_1^{(i)}}}^T(3\vb{x}{\vb*{x}}^T+2T^T T)\vb*{a_1^{(i)}}+{\vb*{a_2^{(i)}}}^T(3\vb{x}{\vb*{x}}^T+2T^T T)\vb*{a_2^{(i)}}\bigg]\\
			&-\frac{1}{81}\bigg[{\vb*{a_1^{(i)}}}^T(3\vb{x}{\vb*{x}}^T+2T^T T)\vb*{a_2^{(i)}}+{\vb*{a_2^{(i)}}}^T(3\vb{x}{\vb*{x}}^T+2T^T T)\vb*{a_1^{(i)}}\bigg]
		\end{split}
		\label{rho_chi_min_1}
	\end{align}
	Thus, the problem of minimization of $\normsq{\rho_{AB}^{3\otimes 3}-\chi_{AB}^{3\otimes 3}}_2$ over all classical states is reduced to find the minimum value of the expression $\normsq{\rho_{AB}^{3\otimes 3}-\chi_{AB}^{3\otimes 3}}_2$ over $\vb*{a_1^{(i)}}$ and $\vb*{a_2^{(i)}}$.
	
	This minimum can be obtained when the vectors $\vb*{a_1^{(i)}}$ and $\vb*{a_2^{(i)}}$ are chosen as the eigenvectors of the $8\times 8$ matrix $3\vb{x}{\vb*{x}}^T+2T^T T$ corresponding to the two largest eigenvalues, denoted by $\lambda_{max}^{(1)}$ and $\lambda_{max}^{(2)}$ respectively, satisfying $\lambda_{max}^{(1)}\geq \lambda_{max}^{(2)}$. Therefore, the minimum value is given by
	\begin{align}
		\begin{split}
			\min_{\chi_{AB}^{3\otimes 3}\in C}{\normsq{\rho_{AB}^{3\otimes 3}-\chi_{AB}^{3\otimes 3}}_2}
			=&\frac{2}{81}\bigg[3\normsq{\vb*{x}}_2+2\normsq{T}_2-\left(\lambda_{max}^{(1)}+\lambda_{max}^{(2)}\right)\bigg]
		\end{split}
		\label{rho_chi_min_2}
	\end{align} 
	Thus, (\ref{rho_chi_min_2}) gives the required formula for the geometric discord for $3\otimes 3$ system.

	\section{Detailed calculations of geometric discord of the $3\otimes 3$ dimensional state $\rho_{\alpha}$}
	Recall the density operator $\rho_{\alpha}$ as
	\begin{align}
		\rho_{\alpha}=\frac{1}{21}\begin{pmatrix}
			2 & 0 & 0 & 0 & 2 & 0 & 0 & 0 & 2 \\
			0 & a & 0 & 0 & 0 & 0 & 0 & 0 & 0 \\
			0 & 0 & 5-a & 0 & 0 & 0 & 0 & 0 & 0 \\
			0 & 0 & 0 & 5-a & 0 & 0 & 0 & 0 & 0 \\
			2 & 0 & 0 & 0 & 2 & 0 & 0 & 0 & 2 \\
			0 & 0 & 0 & 0 & 0 & a & 0 & 0 & 0 \\
			0 & 0 & 0 & 0 & 0 & 0 & a & 0 & 0 \\
			0 & 0 & 0 & 0 & 0 & 0 & 0 & 5-a & 0 \\
			2 & 0 & 0 & 0 & 2 & 0 & 0 & 0 & 2
		\end{pmatrix}
	\end{align}
	In this case we get $\vb*{x}=(0~0~0~0~0~0~0~0)$ and 
	\begin{align}
		T=\begin{pmatrix}
			\frac{4}{21} & 0 & 0 & 0 & 0 & 0 & 0 & 0 \\
			0 & \frac{4}{21} & 0 & 0 & 0 & 0 & 0 & 0 \\
			0 & 0 & \frac{4}{21} & 0 & 0 & 0 & 0 & 0 \\
			0 & 0 & 0 & -\frac{4}{21} & 0 & 0 & 0 & 0 \\
			0 & 0 & 0 & 0 & -\frac{4}{21} & 0 & 0 & 0 \\
			0 & 0 & 0 & 0 & 0 & -\frac{4}{21} & 0 & 0 \\
			0 & 0 & 0 & 0 & 0 & 0 &  -\frac{1}{21} & \frac{-5+2a}{7\sqrt{3}} \\
			0 & 0 & 0 & 0 & 0 & 0 & \frac{5-2a}{7\sqrt{3}}& -\frac{1}{21}
		\end{pmatrix}
	\end{align}
	Therefore, we have
	\begin{align}
		\begin{split}
			\normsq{x}&=0\\
			\normsq{T}&=\frac{8}{441} (31 + 3 (-5 + a) a)
		\end{split}
	\end{align}
	Eigenvalues of $3\vb{x}{\vb*{x}}^T+2T^T T$ are $\frac{32}{441}$ with multiplicity 6 and $\frac{8}{441}(19+3(-5+a)a)$ with multiplicity 2.\\
	(i) \textbf{Separable state:} The state  $\rho_{\alpha}$ given in (\ref{gd_sep_rho_alpha}) is separable when $\alpha\in [2,3]$. For this state, we have $\lambda_{max}^{(1)}=\lambda_{max}^{(2)}=\frac{32}{441}$. So we get 
	\begin{align}
		\begin{split}
			D_G^{(3,3)}(\rho_{\alpha})=\frac{32}{11907}\bigg[\left(\alpha-\frac{5}{2}\right)^2+\frac{11}{4}\bigg], ~2\leq \alpha\leq 3
		\end{split}
		\label{our_separable_alpha_d_proof}
	\end{align}
	
	(ii) \textbf{PPT Entangled state:} The state $\rho_{\alpha}$ given in (\ref{gd_pptes_rho_alpha}) is PPTES for $3< \alpha \leq 4$. In this case, the eigenvalues $\lambda_{max}^{(1)}$ and $\lambda_{max}^{(2)}$ are given by
	\begin{align}
		\lambda_{max}^{(1)}=\lambda_{max}^{(2)}=	
		\begin{cases}
			\frac{32}{441}& 3 < \alpha \leq 3.618\\
			\frac{8}{441}(19+3a(-5+a)) & 3.618 <\alpha\leq 4
		\end{cases}
	\end{align}
	Therefore, the geometric discord of PPTES $\rho_{\alpha}$ can be expressed as
	\begin{align}
		D^{(3,3)}_G(\rho_{\alpha})=
		\begin{cases}
			\frac{32}{11907}(\alpha^2-5\alpha+9)& 3< \alpha \leq 3.618\\
			\frac{128}{11907} & 3.618<\alpha\leq 4
		\end{cases}
		\label{our_PPTES_alpha_d_proof}
	\end{align}
	
	(iii) \textbf{NPT Entangled state:} The state $\rho_{\alpha}$ given in (\ref{gd_nptes_rho_alpha}) is NPTES in the range $4<\alpha\leq 5$. In this case, $\lambda_{max}^{(1)}=\lambda_{max}^{(2)}=\frac{8}{441}(19+3(-5+a)a)$. Thus, the geometric discord can be calculated as
	\begin{align}
		\begin{split}
			D^{(3,3)}_G(\rho_{\alpha})&=\frac{128}{11907}
		\end{split}
		\label{our_NPTES_alpha_d_proof}
	\end{align}

					\end{appendices}
					\end{document}